\documentclass[
 aps,%
 prb,
 reprint,%
 groupedaddress,%
 superscriptaddress,%
 showpacs,%
 letterpaper,%
 footinbib,%
 noeprint%
]{revtex4-1}

\RequirePackage{amsmath,amsfonts,amssymb,bm,braket}
\RequirePackage{graphicx}
\RequirePackage[usenames,dvipsnames]{xcolor}

\RequirePackage{hyperref}
\hypersetup{%
  breaklinks = {true},
  citecolor = {blue},
  colorlinks = {true},
  linkcolor = {red},
  pdfauthor = {Ridolfi et al},
  pdfcreator = {\LaTeX\ and \flqq hyperref\frqq},
}

\renewcommand{\vec}[1]{\text{$\boldsymbol{#1}$}}

\newcommand{\Fref}[1]{Fig.~\ref{#1}}

\renewcommand{\eqref}[1]{(\ref{#1})}
\newcommand{\Eqref}[1]{Eq.~(\ref{#1})}
\newcommand{\Eqsref}[1]{Eqs.~(\ref{#1})}
\newcommand{\Ref}{Ref.}
\newcommand{\Refs}{Refs.}

\newcommand{\be}{\begin{equation}}
\newcommand{\ee}{  \end{equation}}
\newcommand{\ba}{\begin{eqnarray}}
\newcommand{\ea}{  \end{eqnarray}}
\newcommand{\ve}{\varepsilon}
\newcommand{\bk}{\vec{k}}
\newcommand{\bq}{\vec{q}}
\newcommand{\br}{\vec{r}}
\renewcommand{\Re}{\operatorname{Re}}
\newcommand{\teq}{{\,=\,}}
\newcommand{\tequiv}{{\,\equiv\,}}
\newcommand{\tsim}{{\,\sim\,}}
\newcommand{\tsimeq}{{\,\simeq\,}}
\newcommand{\tapprox}{{\,\approx\,}}

\newcommand{\tminus}{{\,-\,}}
\newcommand{\tpm}{{\,\pm\,}}

\graphicspath{ {./Figs/} } 

\begin{document}

\title{Excitonic structure of the optical conductivity in 
MoS$_2$ monolayers}

\author{Emilia Ridolfi}
\affiliation{%
  Centre for Advanced 2D Materials, National University of Singapore,
  6 Science Drive 2, Singapore 117546}

\author{Caio H. Lewenkopf}
\affiliation{%
  Instituto de F\'{\i}sica,
  Universidade Federal Fluminense, 24210-346 Niter\'oi, Brazil}

\author{Vitor~M.~Pereira}
\email[Corresponding author: ]{vpereira@nus.edu.sg}
\affiliation{%
  Department of Physics, National University of Singapore,
  2 Science Drive 3, Singapore 117542}
\affiliation{%
  Centre for Advanced 2D Materials, National University of Singapore,
  6 Science Drive 2, Singapore 117546}
 
\date{\today}

\begin{abstract}
We investigate the excitonic spectrum of MoS$_2$ monolayers and calculate its 
optical absorption properties over a wide range of energies. Our approach takes 
into account the anomalous screening in two dimensions and the presence of a 
substrate, both cast by a suitable effective Keldysh potential. We solve the 
Bethe-Salpeter equation using as a basis a Slater-Koster tight-binding model 
parameterized to fit the \emph{ab initio} MoS$_2$ band structure calculations. 
The resulting optical conductivity is in good quantitative agreement with 
existing measurements up to ultraviolet energies. We establish that the 
electronic contributions to the C excitons arise not from states at the 
$\Gamma$ point, but from a set of $\bk$-points over extended portions of 
the Brillouin zone. Our results reinforce the advantages of approaches based on 
effective models to expeditiously explore the properties and tunability of 
excitons in TMD systems. 
\end{abstract}


\maketitle

\section{Introduction}

The widespread availability of bulk trigonal molybdenum disulfide (MoS$_2$) has 
made this material  one of the most widely studied transition metal 
dichalcogenides (TMDs) --- especially at the strict monolayer thickness ---, and 
has propelled MoS$_2$ to one of the most prominent members of the family of 
semiconducting two-dimensional materials beyond graphene \cite{wang2012, 
Geim2013, Butler2013, Choi2017, WangReview2017}. 
In parallel with the interest and continued advances in optimizing sample 
production and transport characteristics, MoS$_2$ and other closely related TMDs 
are of great appeal for optoelectronic applications \cite{wang2012, Choi2017, 
WangReview2017}. This has sustained intensive research to understand the 
processes that govern the electronic response of these crystals to 
light. Much progress has been made theoretically \cite{Komsa2012, Malic2014, 
Molina2013, Berchelbach2013, Qiu2013, Pedersen2014, Zhang2014, Wu2015, Jose2016, 
Nuno2017, Maxim2017, Malte2016, wang-substrate2017, Attaccalite2014, Glazov2017, 
Rhim2015, Wang2015-dft} and experimentally \cite{Mak2010, Splendiani2010, 
LiRao2013, Zhang_exciton2014, Zhang2014, Li2014, Kim2014, Miwa2015, Hill2015, 
Klots2014, Rigosi2016, Aleit2016, Chiu2015, Malte2016, Chiu2015, Kumar2013, 
Clark2014, Mishina2015, Pedersen2015, Rostami2016, Woodward2017, Xiaobo2014, 
Zhang2015-TwoPhotons} in both understanding fundamental properties and exploring 
the potential practical uses of these materials in devices. 

As in nearly all strictly two-dimensional materials, MoS$_2$ monolayers have a 
highly tunable carrier density \cite{Butler2013, Liu2017, Saito2016, Wang2015} 
and are amenable to having a number of properties tailored on-demand by 
different external procedures, including the customization of the optical 
band-gap \cite{Raja2017}. 
It was early recognized that the intrinsic two-dimensionality and 
semiconducting character of TMDs bring about enhanced Coulomb interactions 
which, not only renormalize the electronic band structure with quantitative 
consequences for all derived single-particle processes, but also give rise to 
the strongest excitonic effects seen to date in the optical response of 
semiconductors \cite{Nakajima1980, Chirolli2017, Crommie2014}.  
With binding energies as high as $0.22{\,-\,}1.1$\,eV  
\cite{Komsa2012, Malic2014, Berchelbach2013, Qiu2013, Pedersen2014, Wu2015, 
Jose2016, Nuno2017, 
Zhang_exciton2014, Zhang2014, Hill2015, Klots2014, Rigosi2016, Chiu2015} 
(depending on the strength of the interaction due to the sensitivity to their 
environment)  and carrying a large spectral weight \cite{Qiu2013, Li2014, 
Mak2010}, these two-particle excitations determine and dominate the optical 
response of TMD materials. As a result, excitons are now understood as a critical 
ingredient in any reliable theory and model of the optical properties of TMDs, to the 
extent that any theory that does not account for excitonic effects fails to 
capture even the most basic qualitative features of the optical gap and/or 
spectral weight distribution.

\begin{figure*}[tb]
\centering
\includegraphics[width=0.75\columnwidth]{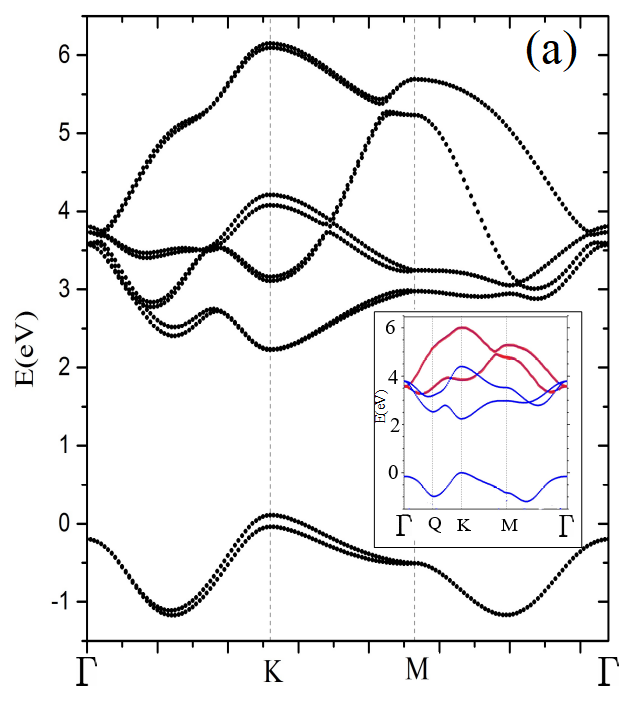} 
\qquad
\includegraphics[width=1.05\columnwidth]{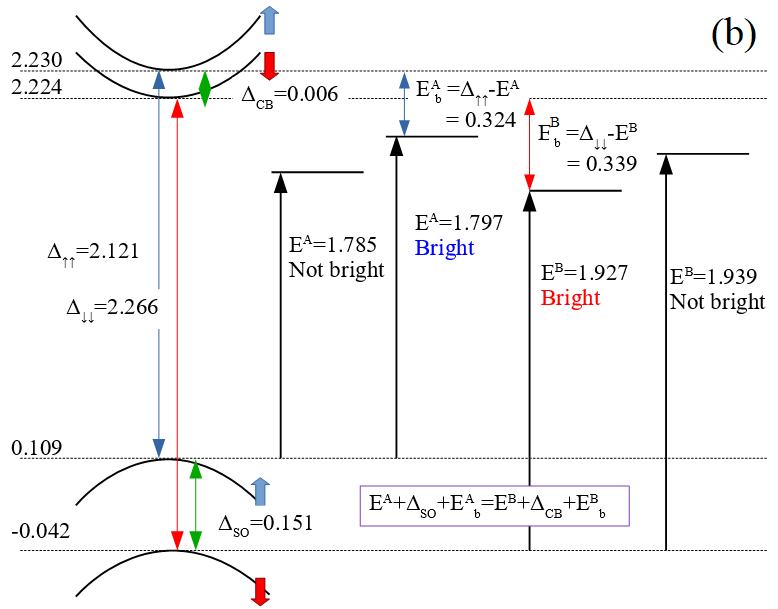}
\caption{
(a) The reference band structure (including SO coupling) of the MoS$_{2}$ 
monolayer according to the Slater-Koster parameterization discussed in the text 
and reported earlier in \Ref~\onlinecite{Ridolfi2015}. The full model 
Hamiltonian can be decoupled into odd and even blocks according to symmetry 
with respect to the Mo plane ($z$-reflection symmetry). The inset shows the 
band structure (without SO coupling) with even (odd) bands highlighted 
in blue (red) color.
(b) Schematic (not to scale) representation of the energy dispersion in the 
vicinity of the fundamental gap and various energy scales related to 
our excitonic spectrum, including the optical bandgaps derived from each 
exciton series ($E^{A,B}$) and the exciton binding energies ($E_b^{A,B}$). This 
panel deliberately exaggerates the 6\,meV splitting of the conduction bands due 
to SO coupling, in order to make clear the origin of two different 
excitation energies for each series in our results [e.g., $E^A$ bright and not 
bright in \Fref{fig:spectrum}(a)]. The experimental A and B peaks correspond to 
$E^A$ (bright) and $E^B$ (bright), respectively, and the represented quantities 
are related through $E^A+\Delta_{SOC}+E^A_b\teq E^B+\Delta_{CB}+E^B_b$.
}
\label{fig:bands}
\end{figure*}

Monolayers of semiconducting TMDs are also interesting due to a number of other 
fundamental and unique features of their electronic structure that can broaden 
their range of applicability in optoelectronics. For example, the strong 
spin-orbit (SO) coupling splits the valence bands at the $K$ point by a large 
amount ($\Delta_{SOC}$) which generates two families of excitons \cite{Miwa2015} 
[see A and B excitons in \Fref{fig:bands}(b)] and allows the selective 
excitation of electrons with predefined spin polarization \cite{Xiao2012}. 
Moreover, the non-zero Berry curvature offers a number of opportunities to 
explore applications related to the non-trivial topological nature of electronic 
states near the band edges. These include the facile injection of 
valley-polarized carriers by optical pumping \cite{Mak2012, Zheng2012, 
Sallen2012, Cao2012}, the ability to control spin and valley populations 
simultaneously as a result of the spin-valley locking \cite{Bawden2016}, or the 
anomalous splitting of bound excitonic levels due to a pseudo spin-orbit 
coupling of topological origin \cite{Zhou2015, Srivastava2015}. 

In view of this, the development of reliable models with enough flexibility to 
allow the prediction of the optical response of TMDs in different experimental 
settings is clearly of high interest. Ideally, one wishes a scheme that augments 
the reach and expediency of accurate and unbiased first-principles calculations 
of the full excitonic spectrum. The latter are notoriously demanding from the 
numerical point of view and, in addition, are particularly onerous for 
2D materials when reasonable convergence is required \cite{Qiu2013, Louie2000}. 
It then becomes prohibitive to rely only on these approaches to scan a 
potentially large scope of modifications (structural, chemical, electronic) that 
can be of interest to tailor the material's intrinsic response for specific 
purposes. 

In this paper our focus are single-layer systems. Henceforth, except if 
explicitly emphasized otherwise, we shall refer to the MoS$_2$ monolayer as 
simply MoS$_2$. The approach that we describe here begins with an accurate 
Slater-Koster (SK) tight-binding parameterization of the target band structure 
\cite{Ridolfi2015}. The model parameters are benchmarked against information 
from first-principles calculations and experiments to describe the most 
important spectral features of MoS$_2$, such as the correct energies and orbital 
content of the low lying conduction and valence bands at the critical points in 
the Brillouin zone (BZ). 
We are able to reproduce optical absorption spectra obtained experimentally
with quantitative accuracy in both frequency and absolute magnitude.
As expected, our calculations agree with previous theoretical works  
\cite{Ramasubramanian2012, Qiu2013, Berchelbach2013, Klots2014, Komsa2012} 
that provide a good description of the strongly localized $A$ and $B$ excitons.
By using a large sampling of $\bk$ points in the Brillouin zone we are also 
able to study the so-called C-exciton \cite{Qiu2013} and establish its nature, a 
subject under debate in the literature 
\cite{Qiu2013,Klots2014,Kim2014,Aleit2016}.

The remainder of this paper is organized as follows: In 
Sec.~\ref{sec:state-of-the-art} we discuss the state-of-the-art experimental and 
theoretical work on the optical response of TMD monolayers. 
Section~\ref{sec:theory} presents our solution of the BSE using a SK Hamiltonian 
optimized for MoS$_2$ monolayers and its use in calculating the optical 
conductivity in linear response. The results, with focus on the nature of the 
resonant C excitons, are discussed in Sec.~\ref{sec:results}. Finally, a summary 
of our main findings is presented in Sec.~\ref{sec:conclusions}. The paper also 
includes one appendix that addresses technical issues, such as the choice of the 
number of bands taken in the calculation, an analysis of the spin-orbit effects 
in the optical response, and a comparison between the energy spectrum obtained 
with and without the Coulomb interaction.

\section{Excitons in the optical response of MoS$_2$} 
\label{sec:state-of-the-art}
 
The optical conductivity, absorption, and reflectance spectra of few and 
single-layer MoS$_{2}$ has been extensively studied in recent experiments 
\cite{Mak2010, Splendiani2010, LiRao2013, Zhang_exciton2014, Zhang2014, Li2014, 
Kim2014, Miwa2015, Hill2015, Klots2014, Rigosi2016, Aleit2016}. 
It is well 
established that the onset of optical absorption in clean, undoped monolayers 
occurs at $1.8\tpm 0.1$\,eV,  as measured spectroscopically  by 
reflection and photoluminescence \cite{Mak2010, Splendiani2010, Zhang2014, 
Li2014, Klots2014, Rigosi2016}, absorption \cite{Zhang_exciton2014, Li2014, 
Kim2014, Aleit2016}, photoemission \cite{Miwa2015}, and second-harmonic 
analysis \cite{Kumar2013, Pedersen2015, Rostami2016}.

The absorption threshold is characterized by two peaks separated by $145\tpm 
4$\,meV \cite{Miwa2015}, associated with the two families of 
excitons (A and B) derived from transitions between the spin-split valence and 
conduction bands.
Studies of angle-resolved photoemission spectroscopy \cite{Zhang2014}, as well 
as X-ray photoemission and scanning tunneling microscopy/spectroscopy 
\cite{Chiu2015}, show that the single-particle gap lies within $2.15-2.35$\,eV, 
thereby placing the binding energies of the lowest A exciton at $0.22-0.42$\,eV 
(we note that the non-negligible temperature dependence of the absorption peaks 
is an important factor when extracting binding energies and identifying 
experimental variability \cite{Malte2016, Molina2016,Tongay2012, 
Kioseoglou2016}).
Such large binding energy values imply unusually small exciton radii, 
typically on the order of $\tsim 5$\,\AA \cite{Wu2015}, but still 
of the Wannier-Mott type. 

Theoretically, the solution of the BSE from first principles on the basis of 
GW-corrected electronic states captures accurately the experimental behavior 
related to the A and B excitons \cite{Ramasubramanian2012, Qiu2013, 
Berchelbach2013, Klots2014, Komsa2012}. Yet, they also reveal the numerical 
challenges intrinsic to a full \emph{ab-initio} approach to this problem in 2D 
systems, which is particularly demanding 
\footnote{For example, \Ref~\onlinecite{Qiu2013} attributes the peak C 
to transitions near, but not directly at, the $\Gamma$ point, which require a 
fine sampling with $300^2$ $\bk$ points, and at least $56$ bands in the 
underlying GW calculation. The authors also use local field effects to include 
the interaction over different BZs.}
in terms of convergence at both the stage of the GW single-particle corrections 
and the subsequent solution of the BSE \cite{Ramasubramanian2012, Qiu2013, 
Berchelbach2013, Klots2014, Louie2000, Cudazzo2011}.
Effective models, on the other hand, must cope with the non-Coulomb form of the 
screened potential which is essential to capture the correct bound exciton 
series \cite{Chernikov2014, Rodin2014, Keldysh1973}, but prevents closed-form 
analytical results for the binding energies or wave functions.
$\bk{\cdot}\bm{p}$ models that describe the conduction and valence valleys in 
terms of a massive Dirac equation adapted to MoS$_2$ \cite{Berchelbach2013, 
Zhang_exciton2014, Jose2016, Xiao2012}  have been able to capture the bound 
excitonic series \cite{Berchelbach2013, Zhang_exciton2014, Jose2016}, the 
momentum dispersion of the excitonic spectrum \cite{Wu2015}, and the 
excitonic contributions to the optical conductivity \cite{Zhang_exciton2014, 
Xiao2012, Nuno2017}. 
A prevalent characteristic of studies based on these  models is their focus on 
specific features, most notably the energy spectrum itself which is 
non-hydrogenic \cite{Chernikov2014, Maxim2017, Srivastava2015} and had not been 
correctly described until recently. Whereas such a restricted analysis is an 
implicit requirement of effective mass approaches, it is not a limitation for 
models based on TB, which can describe the entire BZ and large energy ranges of 
relevance for experiments and applications, provided the starting Hamiltonian 
gives a quantitatively accurate and qualitatively faithful description of the 
single-particle states.
Parametrized models, both of the SK type as well as simpler, orbital 
non-specific TB Hamiltonians, have also been employed with different levels of 
accuracy and reproducibility of experiments \cite{Malic2014, Pedersen2014, 
Nuno2017, Wu2015}: some report results on restricted energy ranges around the A 
and B peaks \cite{Malic2014, Wu2015}, others resort to TB models with a large 
number of fitting parameters (e.g., $>28$) \cite{Pedersen2014, Wu2015} or 
include only a basis of $d$ orbitals \cite{Nuno2017, Wu2015} (the orbital 
character becomes relevant away from the $K$ point; for example, 
\Ref~\onlinecite{Qiu2013} identifies the C excitons with states that have both 
Mo $d_{z^2}$ and S $p_{x,y}$ character). Most importantly, the single particle 
band structure of some of these calculations does not capture well the 
GW-corrected band gap \cite{Wu2015} or the dispersion of the upper conduction 
bands \cite{Jose2016}, which is especially relevant factors in the excitonic 
problem. 

We note, finally, that, due to the zero crystal momentum involved in the 
underlying electronic excitations, the excitonic fingerprints in the optical 
properties of bulk MoS$_2$ are qualitatively and quantitatively similar to those 
of the monolayer. This follows from the layered structure of the former which, 
combined with a relatively weak inter-layer electronic coupling, makes the 
electronic properties of the bulk strongly two-dimensional in character. Not 
surprisingly, and despite the different screening environment, excitons in bulk 
samples tend to have binding energies and radii similar to those occurring in 
the monolayers \cite{Saigal2016}, and remain mostly localized within one layer 
\cite{Molina2013}. 
Understanding the excitonic physics in the monolayer is therefore key for the 
description of the corresponding physics in the bulk as well.

\section{Theory and methods} 
\label{sec:theory}
\subsection{Exciton states and the BSE}
 
Neutral excitations in crystals, both bound and extended, are well described by 
approximate solutions of the BSE \cite{BSE, BSEsite, DelSole1998, Louie2000, 
Olevano2017}. First principles methods have been widely used to investigate the 
optical properties of insulators and semiconductors in this framework, and 
provide the current standard to tackle the excitonic spectrum of solid-state 
materials \cite{Louie2000}. Since Coulomb interactions and screening are the 
essence of the exciton problem, these have to be properly handled in a 
consistent way to establish even the ``single-particle'' ground-state of the 
system (i.e., its one-particle band structure). This need to self-consistently 
account for quasiparticle corrections in addition to solving the BSE proper 
constitutes a notable  challenge both in terms of implementation and in 
computational time. Thorough and converged first-principle calculations of the 
excitonic spectrum and related observables in MoS$_2$ have, as a result, been 
typically few and far between \cite{Ramasubramanian2012, Qiu2013, 
Berchelbach2013, Klots2014}.

Since the excitonic physics is essential to describe the optical response of 
semiconducting TMDs, and in view of the current need for accurate, yet 
expedite, methods to tackle these properties, we solve the BSE and calculate the 
optical conductivity using an orthogonal SK TB Hamiltonian parameterized to 
describe the MoS$_2$ band structure. The atomic orbital basis comprises the 
three $p$ valence orbitals in each S plus the five $d$ orbitals in each Mo 
within the trigonal unit cell, giving a total of 11 atomic orbitals. The 
construction of the Hamiltonian and optimization of its SK parameters has been 
presented in detail elsewhere \cite{Ridolfi2015}. In brief, its main 
characteristics are: (i) only $14$ fitting parameters, (ii) the correct band gap 
of $2.115$\,eV at the $K$ point, (iii) the spin splitting of the VB at the $K$ 
point by $150$\,meV, (iv) the effective masses and positions of the conduction 
and valence bands at $K$, $\Gamma$ and at the so-called $Q$ point. The TB 
parameters for the SO coupling have been chosen to match $\Delta_{SOC}$ with the 
experimental energy difference between A and B peaks \cite{Miwa2015}. 
The associated band structure is reproduced in \Fref{fig:bands}(a), 
reflecting the insulating ground state of a pristine monolayer.
The Bloch states, $\psi_{n\bk}(\br)$, derived from this Hamiltonian are taken as 
a good approximation to the eigenstates of the crystal Hamiltonian,
\begin{equation}
  \hat{H}\, \psi_{n\bk}(\vec{r}) = \ve_{n\bk}\psi_{n\bk}(\vec{r}),
\end{equation}
where
\begin{equation}
  \psi_{n\bk}(\vec{r}) 
  = \frac{1}{\sqrt{N_c}}\sum_{\vec{R}}e^{i\bk\cdot\vec{ R } }  
  \sum_{\alpha}C_{\alpha,\bk}^{n}\phi_{\alpha}(\vec{r}-\vec{R} -\vec{t}_\alpha)
  \label{eq:bloch}.
\end{equation}
The lattice vector $\vec{R}$ runs over all $N_c$ unit cells of the crystal, $n$ 
is the band index, $\alpha$ denotes the orbitals, and $\vec{t}_\alpha$ 
corresponds to the position in the unit cell of the atoms at which the orbitals 
are centered. Both the band $n$ and orbital $\alpha$ indices run over the 
same interval  $[1,N]$, where $N\teq 22$ ($11{\times}2$, with spin) is the 
dimension of the orbital basis considered in the SK Hamiltonian.  
These Bloch states are used to set up the BSE in the Tamm-Dancoff approximation 
(TDA) by introducing a basis of two-particle excitations of the Fermi sea, 
$\ket{\text{FS}} \tequiv \prod_{v\bk} 
a_{v\vec{k}}^{\dagger}\ket{0}$, 
\begin{equation}
  \ket{v\bk \rightarrow c\bk} \equiv a^\dagger_{c\bk}a^{}_{v\bk} \ket{\text{FS}}.
  \label{eq:basis}
\end{equation}
The latter are used to express the exciton states 
\begin{equation}
  \ket{M} = \sum_{c,v,\bk}A_{cv\bk}^M \ket{v\bk\rightarrow c\bk},
  \label{eq:expansion}
\end{equation}
with energy $E_{M}$, where $M$ labels the excitonic modes.

Note that these definitions implicitly restrict the exciton momentum to zero; 
this is sufficient to capture all first order optical processes, as the excitons 
with zero momentum are the  optically bright ones. Hence, we restrict our 
discussion to this subspace only. Furthermore, we employ the traditional 
notation ``$c/v$'' to designate conduction and valence bands in order to 
emphasize that we shall be working at zero temperature. It is also instructive 
to note at this point that, since $c\in[1,N_{c}]$, and $v\in[1,N_{v}]$ where 
$N_{c}$/$N_{v}$ is the number of conduction/valence bands, the dimension of the 
vector space spanned by the basis states in \Eqref{eq:basis} is $N_\text{tot} 
\tequiv N_k^2 \times N_{c} \times N_{v}$, where $N_k^2$ represents the total 
number of points sampled in the BZ. The number of bands and the size of the 
sampling in $\bk$ points is one of the critical limiting factors in 
calculations of the two-particle spectrum, even within a parameterized TB 
framework. 

Replacing \eqref{eq:expansion} in the Schr\"odinger equation that includes the 
many-body Coulomb interaction yields the reduced eigenproblem 
\cite{Grosso,Pedersen2014,Wu2015}
\begin{equation}
  E_{cv\bk}A_{cv\bk}^{M} + 
  \frac{1}{V} \sum_{c'v'\bk'} W_{cv\bk,c'v'\bk'}A_{c'v'\bk'}^M
  = E_M A_{cv\bk}^M .
  \label{eq:BSE}
\end{equation}
Here, $E_{cv\bk} \tequiv \ve_{c\bk}-\ve_{v\bk}$ is the energy difference between 
the $c$ and $v$ bands  at $\bk$,  $V \tequiv A_{c} N_k^2$ is the total area of 
the crystal ($A_c\teq \sqrt{3}a^{2}/2$ with $a \tsimeq 3.16$\,\AA{} the lattice 
constant) and $W_{cv\bk,c'v' \vec{k'} } \tequiv \bra{v \bk\rightarrow c\bk} 
\hat{U} \ket{v' \vec{k'} \rightarrow c' \vec{k'}}$ represents the matrix element 
of the many-body Coulomb potential, $\hat{U}$, between two particle-hole 
excitations. In the TDA, there are two contributions to this matrix element: a 
direct and an exchange term
\footnote{For the derivation of the BSE in details on the computation of $W$ see,
for instance, Chapters VII.1. and IV.4-5 of \Ref~\onlinecite{Grosso}.}.
As pointed out earlier \cite{Wu2015}, in an orthogonal basis and in our 
approximation where the Coulomb interaction is independent of the orbital 
character of the states involved, the exchange term does not contribute for 
zero-momentum excitons. As a result, only the direct Coulomb matrix element 
remains, which can be expressed simply as \cite{Pedersen2014,Wu2015}
\begin{equation}
  \label{eq:direct-W}
  W_{cv\bk,c'v'\bk'}^{(d)}  =  
  u(\bk-\bk') \, I_{c'\bk',c\bk}^{*}I_{v'\bk',v\bk},
\end{equation}
where $u(\bm{q})$ is the Fourier transform of the \emph{screened} Coulomb 
potential \cite{Louie2000}. The orthogonality assumed in defining our SK basis 
\cite{Ridolfi2015} allows one to express the overlap integrals $I_{a\bk',b\bk}$ 
in terms of the expansion coefficients of the Bloch states \eqref{eq:bloch} as
\begin{equation}
  I_{a\bk',b\vec{ k}} =
  \sum_{\alpha}C_{\alpha\bk'}^{a\,*} \, C_{\alpha\bk}^{b}.
\end{equation}
Since the $C_{\alpha\bk}^{b}$ are obtained from the numerical eigenvectors of 
the Bloch Hamiltonian \eqref{eq:bloch}, it is important to ensure a consistent 
choice of phase because the $I_{a\bk',b\vec{ k}}$ are not gauge-invariant 
quantities.  We chose to require the sum of the basis-set coefficients of the 
wave function $\rho_{\bk}^{n}\teq \sum_{\alpha}C_{\alpha \bk}^{n}$ to be real, 
as suggested in \Ref~\onlinecite{Louie2000}.

\subsection{Screened Coulomb interaction}

An accurate approximation to describe the screened Coulomb interaction in 
\Eqref{eq:direct-W} is essential for a realistic description of the excitonic 
spectrum \cite{Louie2000}. Early attempts to theoretically describe the exciton 
series in MoS$_2$ and related 2D materials provide a good example of this 
stringent requirement, since, by simplistically using the bare Coulomb form of 
the potential, one fails to capture the non-Rydberg level structure observed 
experimentally \cite{Chernikov2014, Maxim2017, Srivastava2015}. The distinct 
series of bound exciton levels in MoS$_2$ is due to  both its pseudospin degree 
of freedom \cite{Maxim2017} and the modified electrostatic interaction in 
strictly 2D electronic systems which, in Fourier space, acquires the form 
\cite{Chernikov2014, Rodin2014}
\begin{equation}
  \label{eq:potential-transform}
  u(\vec{q})=-\frac{e^{2}}{2\epsilon_{0}\epsilon_d\,q\,\kappa(q)},
  \quad
  \kappa(q) \equiv 1 + r_0q,
\end{equation}
where $r_0$ defines the 2D polarizability of the electronic system  
\cite{Cudazzo2011, Chernikov2014, Rodin2014} and $\epsilon_d$ captures the 
static, uniform screening due to the top and bottom media surrounding the 
MoS$_2$ monolayer \cite{Zhang2014}.
We assume a MoS$_{2}$ monolayer of effective thickness $d$ and effective 
dielectric constant $\epsilon_{2}$ sandwiched between materials with dielectric 
constants  $\epsilon_{1}$ and  $\epsilon_{3}$. The environment dielectric 
constant is thus $\epsilon_{d}\teq (\epsilon_{1}+\epsilon_{3})/2$. 
The potential \eqref{eq:potential-transform} has precisely the form derived by 
Keldysh for a thin metallic film, in which case the parameter $r_0$ enters as 
the film thickness \cite{Keldysh1973}. The explicit $q$-dependence in the 
dielectric function $\kappa(q)$ due to many-body interactions qualitatively 
modifies Coulomb's law in real space which becomes
\begin{equation}
  u(\br) = -\frac{e^2}{8\epsilon_{0}\epsilon_{d}r_{0}}
    \Bigl[H_{0}\Bigl(\frac{r}{r_{0}}\Bigr) 
    -Y_{0}\Bigl(\frac{r}{r_{0}}\Bigr)\Bigr]
  ,\label{eq:potential}
\end{equation}
where $H_{0}$ and $Y_{0}$ are Struve and Bessel functions, respectively. The 
parameter $r_0$ defines a crossover 
length scale separating the long-range decay $\propto 1/r$ from the short-range 
domain characterized by a singularity $\propto \log r$ as $r\to 0$ 
\cite{Cudazzo2011}.
\emph{Ab-initio} studies have confirmed that the Keldysh interaction accurately 
describes the screened potential in MoS$_2$ \cite{Berchelbach2013, Latini2015}. 
Thus, we employ $u(\bq)$ in \Eqref{eq:direct-W} to solve the BSE. 

Only the parameters $r_0$ and $\epsilon_{d}$ remain now to fully specify the 
content of \Eqref{eq:BSE}. They have been reported with a large variation among 
different authors in the recent literature \cite{Zhang2014, Wu2015, Jose2016, 
Berchelbach2013}.
\emph{Ab-initio} calculations of the 2D polarizability find $r_0\tsim 
31.2-41.5$\,\AA ~ in vacuum \cite{Berchelbach2013}. However, it is known that 
the precise energy placement of the exciton level series is sensitive to the 
details of the dielectric environment surrounding the monolayer sample 
\cite{Wu2015, Raja2017, Nuno2017, wang-substrate2017}. Prior estimates for the 
monolayer in the dielectric environment of an air/substrate interface report 
values spanning a relatively wide interval, namely, $r_0\tsim 13.55-57.6$\, \AA 
\cite{Zhang2014, Wu2015, Jose2016, Berchelbach2013}. 
Since we will be referring to the measurements by Li and collaborators 
\cite{Li2014}  as our reference for the experimental optical conductivity, the 
environment's dielectric constant is $\epsilon_d\teq 2.5$, as appropriate for 
the air/silica interface ($\epsilon_1\teq 1$, $\epsilon_3\teq 4$).
In the absence of present \emph{ab-initio} calculations of the corrections to 
the polarizability due to the effect of a silica substrate, we follow the 
estimates put forward in Refs.~\onlinecite{Zhang2014, Wu2015}, which are 
based on the Keldysh-type finite thickness ($d$) model:
\begin{equation}
  \label{eq:r0}
  r_0 = \frac{2\epsilon_{2}^2-\epsilon_{1}^2-\epsilon_{3}^2}{2\epsilon_{2}
(\epsilon_{1}+\epsilon_{3})}\,d
.
\end{equation}
In this expression, $\epsilon_2$ stands for an effective dielectric constant of 
MoS$_2$. The best agreement with the measured exciton binding energies is 
obtained with \cite{Zhang2014} $d\tsimeq6$\,\AA\ and $\epsilon_{2}\tsimeq 12$ 
(the latter matches well the results from first principles calculations of the 
dielectric constant of bulk MoS$_{2}$ \cite{Berchelbach2013}), resulting in 
$r_0\teq 13.55$\,\AA%
\footnote{
The expression $r_{0} \teq \frac{\epsilon_{2}d}{\epsilon_{1}+\epsilon_{3}}$ is 
sometimes reported as the limit of \Eqref{eq:r0} when $\epsilon_{2}  \gg 
\epsilon_{1,3}$.
}.

Having thus specified all its contributions, and even though numerically more 
efficient methods have been proposed recently \cite{Pedersen2014}, we solved the 
BSE by full diagonalization of the eigenvalue problem in \Eqref{eq:BSE}. In view 
of the ranges of energy covered in current experiments, we restricted our base 
states in \Eqref{eq:basis} to include excitations between $N_v \teq  2$ valence 
bands ($1\times 2$ since spin is explicitly included) and $N_c\teq 8$ conduction 
bands. 

Note that, when considering effective models in the $\bk\cdot\vec{p}$ (Dirac) 
approximation it is important to additionally include the pseudospin degree of 
freedom in the treatment of the effective Schr\"odinger equation to properly 
describe the spectrum \cite{Maxim2017}.

\subsection{Optical response}
\label{sec:optical_response}

As the $D_{3h}$ point group determines that all rank 2 tensors are in-plane 
isotropic, in a MoS$_2$ monolayer it is sufficient to consider the diagonal 
component of the optical conductivity $\sigma(\omega)\tequiv\sigma_{xx}(\omega)$ 
which is given in linear response by (dipole approximation, $T\teq 0$) 
\cite{Pedersen2014, Fabio2016}
\begin{equation}
  \Re \sigma(\omega) = \frac{e^{2}\pi}{m^2\hbar\omega V}
  \sum_{M}|\bra{\text{FS}}\hat{P}_x\ket{M}|^{2}\delta(\omega-\omega_{M})
  .
  \label{eq:sigma-def}
\end{equation}
Using \Eqref{eq:expansion} to express $\ket{M}$ we write the total momentum 
operator matrix element as
\begin{equation}
  \bra{\text{FS}}\hat{P}_x\ket{M} = \sum_{cv\bk}A_{cv\bk}^{M} 
  \bra{\text{FS}}\hat{P}_{x} a_{c\bk}^{\dagger}a_{v\bk}^{} \ket{\text{FS}} 
  .
  \label{eq:FS-P-M}
\end{equation}
Expanding the many-body momentum operator in the usual way, $\hat{P}_x \teq  
\sum_{pq} \bra{p}\hat{p}_x\ket{q} a_{p}^{\dagger}a_{q}$, one has
\begin{multline}
  \bra{\text{FS}}\hat{P}_{x} a_{s}^{\dagger}a_{r} \ket{\text{FS}} 
  = \\ \sum_{pq}\bra{p}\hat{p}_{x}\ket{q} \bigl[  
  \delta_{pr}\delta_{qs}f_r(1-f_s) + \delta_{pq}\delta_{sr} f_p f_s     
  \bigr],
\end{multline}
where $f_j$ corresponds to the Fermi-Dirac occupation number of an electron at 
the state $j$. At zero temperature and noting that we are interested in the case 
where $c\ne v$, we write
\begin{align}
  \bra{\text{FS}}\hat{P}_{x} a_{c\bk}^{\dagger}a_{v\bk}^{} \ket{\text{FS}} 
  & =
  \bra{\psi_{v\bk}}\hat{p}_x\ket{\psi_{c\bk}}
  \nonumber \\ 
  & = 
  \frac{m}{\hbar}\bra{\psi_{v\bk}}\nabla_{k_x} \hat{H}(\bk)\ket{\psi_{c\bk}}
  .
  \label{eq:FS-P-aa-FS}
\end{align}
Hence, inserting here the expression for Bloch states given in \Eqref{eq:bloch}, 
the optical conductivity \eqref{eq:sigma-def} becomes
\begin{widetext}
\begin{equation}
  \Re \sigma (\omega) = 
  \frac{e^{2}}{4\hbar}\,\frac{4\pi}{\hbar\omega V}
  \sum_{M} 
  \biggl|
    \sum_{\bk cv}A_{cv\bk}^M \sum_{\alpha \beta} 
      (C_{\alpha \bk}^v)^* C_{\beta \bk}^c 
      \nabla_{k_x} \! \bra{\phi_\alpha} \hat{H}(\bk) \ket{\phi_\beta}
  \biggr|^2
  \delta(\hbar\omega-\hbar\omega_M).
  \label{eq:sigma-final}
\end{equation}
This form makes explicit that the oscillator strength associated with each 
particle-hole excitation involves contributions that depend both on the 
solution of the BSE (through the eigenvector components $A^M_{cv\bk}$) and on 
the effective Hamiltonian in the crystal momentum representation (through the 
components $C_{m\bk}^c$).

The response in the non-interacting approximation (single-particle) is readily 
recovered by noting that, in the absence of particle-hole interaction, the BSE, 
\Eqref{eq:BSE}, is diagonal. In this limit, $A_{cv\bk}^M \to \delta_{cv\bk,M}$, 
$E_M \to E_{cv\bk}$, and 
\Eqref{eq:sigma-final} simplifies to 
\begin{equation}
  \Re \sigma_{\rm sp} (\omega) = 
  \frac{e^{2}}{4\hbar}\,\frac{4\pi}{\hbar \omega V}
  \sum_{cv\bk} 
  \biggl|
    \sum_{\alpha \beta} (C_{\alpha\bk}^v)^* C_{\beta\bk}^c 
      \nabla_{k_x} \! \bra{\phi_\alpha} \hat{H}(\bk) \ket{\phi_\beta}
  \biggr|^2
  \delta(\hbar\omega - E_{cv\bk}).
  \label{eq:sigma-single}
\end{equation}
\end{widetext}

\begin{figure}[tb]
\centering
\includegraphics[width=\columnwidth]{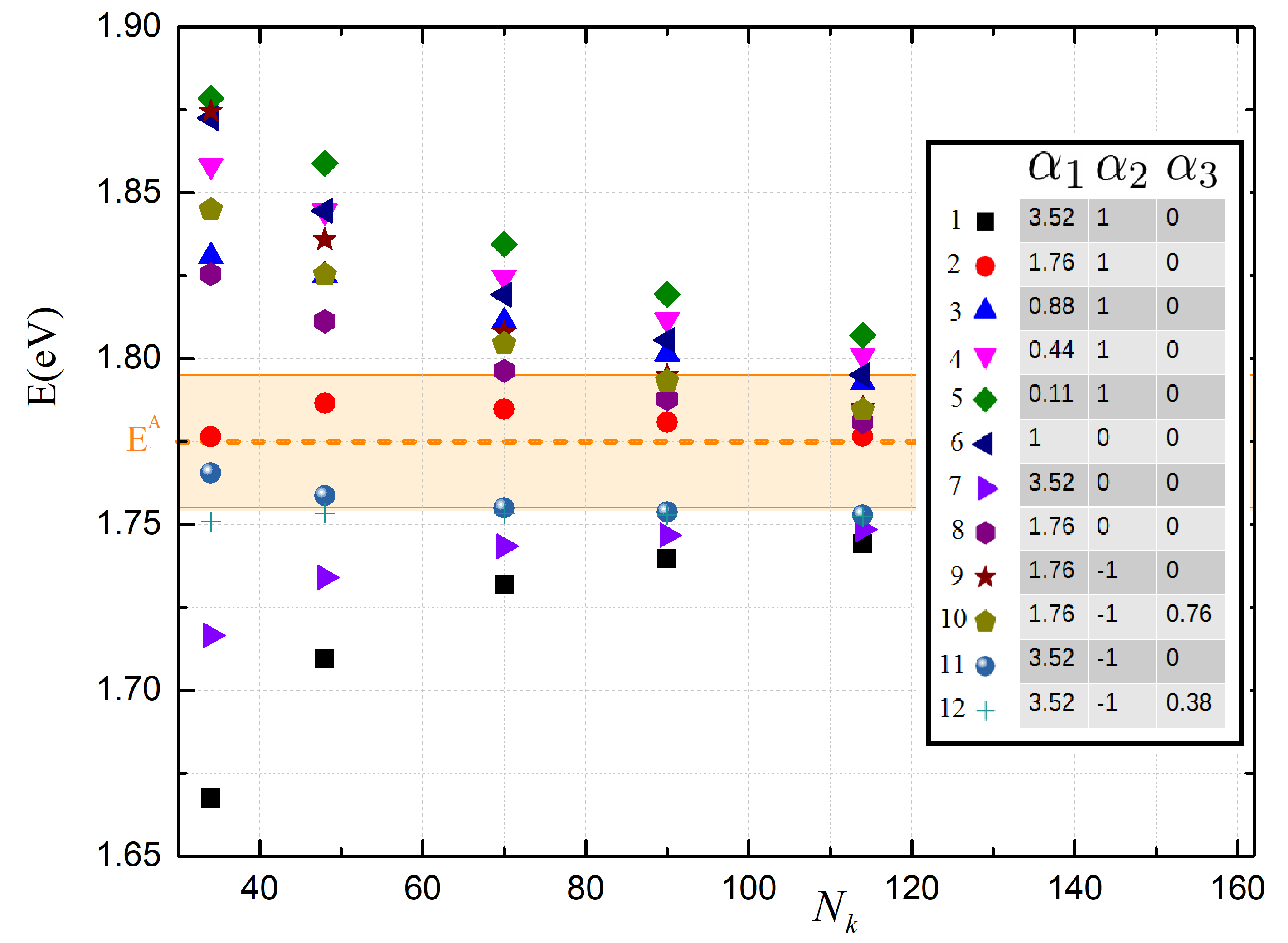}
\caption{
The energy of the lowest eigenvalue of the BSE [$E^A\teq 1.775$ eV, dark, cf. 
\Fref{fig:spectrum}(a)] as a function of the total number of $\bk$ points used 
in the uniform sampling of the Brillouin zone. The horizontal dashed line 
indicates the value obtained from the combined extrapolation of all the curves 
corresponding to the different regularizations. The shaded strip identifies the 
energy interval within $\tpm 20$\,meV of the extrapolated result. 
}
\label{fig:convergence}
\end{figure}

\section{Results}
\label{sec:results}

\subsection{Convergence of the exciton spectrum}

We solve the eigenproblem in \Eqref{eq:BSE} using a uniform sampling of $N_k$ 
points along the directions defined by the reciprocal lattice vectors ${\bf 
b}_{1} \teq  (2\pi/a,\,2\pi/a\sqrt{3})$, ${\bf b}_{2} \teq  
(2\pi/a,-2\pi/a\sqrt{3})$. We place the $\Gamma$ point at the origin, leaving 
$K$ and $M$ at the center of our BZ sampling domain 
[see \Fref{fig:wavefunctions}(a) below]. 
Since the Coulomb interaction is not periodic in the reciprocal space and our 
approach does not include local field terms ({\it i.e.}, Fourier components 
beyond the first BZ), the selection of the sampling domain can have an impact on 
the energies of the bound excitonic levels, especially when these arise from 
transitions at $\bk$ points near BZ boundaries. More specifically, in the 
summation performed in \Eqref{eq:BSE}, $\bk'$ formally runs over the whole 
reciprocal space, which can be expressed by defining $\bk' \tequiv {\bk}'' + 
\bm{Q}$, with ${\bk}'$ spanning the first BZ and $\bm{Q}$ the reciprocal 
lattice. When the excitonic states have wave functions strongly localized near 
the edges of the BZ, one must include $\bm{Q}{\,\neq\,}0$ terms, otherwise one 
misses important contributions from Coulomb matrix elements between $\bk$ states 
closely spaced, but in adjacent Brillouin zones. Since the wave functions 
relevant to our problem are localized in regions around the $\bm{K}$ points in 
the BZ (to be discussed below, see \Fref{fig:wavefunctions}), our choice of the 
$\bk$-domain gives converged results for the excitonic spectrum in agreement 
with the experiment reported in \Ref~\onlinecite{Li2014} by keeping only $\bm{Q} 
\teq 0$ (i.e. by considering all wave vectors and matrix elements within the 
first BZ). Being able to work with this truncation of the Coulomb matrix 
elements without affecting the convergence of the spectrum provides an 
additional improvement in the numerical efficiency of the calculation.

Note, however, that any discrete approach requires the regularization of the 
$\bk$-diagonal matrix elements $W_{cv\bk,c'v'\bk}$ due to the  $\propto 
1/|\bk-\bk'|$ integrable singularity that arises from the long range tail of the 
screened potential [cf. \Eqref{eq:potential-transform}]. As their contributions 
to \Eqref{eq:BSE} become regular in the thermodynamic limit $N_k\gg1$, this 
regularization is well posed in the sense that the specific way it is performed 
has no influence on the calculated observables in that limit. However, we find 
that the regularization strategy \emph{significantly} impacts the rate of 
convergence of the spectrum with $N_k$, to the extent that one can gain an order 
of magnitude reduction in the dimension of the Bethe-Salpeter matrix for a given 
convergence target. This is a point of high practical significance because, in 
an effective SK description such as ours, the number of bands is, by 
construction, the minimal \emph{a priori} required set. The BZ sampling thus 
becomes the limiting factor determining the size and tractability of the 
numerical problem for a given target precision of the calculated spectrum.
Since the singular matrix elements are integrated over the $\bk$-space, the most 
straightforward regularization consists in replacing $u(\bq\teq \bk-\bk'\teq 0)$ 
by its average value over a small enclosing domain $\Delta_\bq$, namely,
\begin{align}
  u(\bq\approx0) & \to \frac{1}{\Delta_\bq} \int_{\Delta_\bq} u(\bq) d\bq 
  \nonumber \\
  & \approx \frac{ca{N_k}}{2\pi} \left[ \alpha_1 
  + \alpha_2\frac{2\pi r_0 }{a{N_k}}
  + \alpha_3 \Bigl(\frac{2\pi r_0 }{a{N_k}}\Bigr)^2 \right]
  .
  \label{eq:regularization}
\end{align} 
Here, $c\tequiv-e^2 / 2 \epsilon_0 \epsilon_d$ and the constants $\alpha_j$ 
depend on the geometry of the averaging domain $\Delta_\bq$ and the truncation 
level of the expansion \eqref{eq:regularization}.

\Fref{fig:convergence} illustrates the different convergence rate of the lowest 
exciton level ($E^A$) as a function of the sampling dimension. For demonstration 
purposes, these data were obtained by solving \Eqref{eq:BSE} with only the two 
bands of each spin nearest to the bandgap (i.e., $N_c\teq N_v\teq 2$). The 
asymptotic value is clearly approached much faster for certain choices of the 
regularization scheme. In particular, one sees that neglecting the leading 
higher order terms in the expansion \eqref{eq:regularization} by having 
$\alpha_1\teq 1$, $\alpha_{2,3}\teq 0$ (parameter set \#$6$, see label in 
\Fref{fig:convergence}) provides a particularly slow convergence%
\footnote{
Having $\alpha_1\teq 1$, $\alpha_{2,3}\teq 0$ in \Eqref{eq:regularization} 
corresponds to leaving the factor $\propto 1/\kappa(q)$ in the potential 
\eqref{eq:potential-transform} outside of the average integral.
}. 
This is not unexpected because $2\pi r_0 / a \tsimeq 26.9$, which is precisely 
of the same magnitude as those values of $N_k$ that are within practical 
numerical reach%
\footnote{
Recall that if $N_k\teq 100$, the full diagonalization of the BSE Hamiltonian 
requires handling a matrix of dimension $N_\text{tot}\times N_\text{tot} \teq  
(10^4 N_c N_v)^2$. For us, with $N_c \teq  8$ and $N_v \teq  2$, that amounts 
to $\tsimeq 2.56 \times 10^{10}$.
}.
On the other hand, the particular $\alpha$ parameter sets \#$2$ and \#$11$ 
(see labels in Fig.~\ref{fig:convergence})  provide much faster convergence to 
the asymptotic value $E^A \teq  1.775$\,eV within a $0.020$\,eV 
precision%
\footnote{
In particular, \#$11$ consists of choosing $q\teq 0$ at the center of a square 
integration domain of side ${2\pi }/{a{N_k}}$ and \#$2$ in placing $q\teq 0$ at 
the corner of the same integration square.
}.
This translates into a binding energy $E_b^A \teq  0.34$\,eV since the 
single-particle gap in our SK band structure parametrization is $E_g \teq  
2.12$\,eV \cite{Ridolfi2015} (this value corresponds to the first dark $A$ 
exciton, while the first bright one has $E^A \teq  1.80$\,eV and binding energy 
$E_b^A \teq  0.32$\,eV).
Henceforth, all our calculations will be presented according to the 
regularization scheme \#$2$. We have verified that it works efficiently for the 
whole spectrum. 

\begin{figure}[tb]
\centering
\includegraphics[width=0.35\columnwidth]{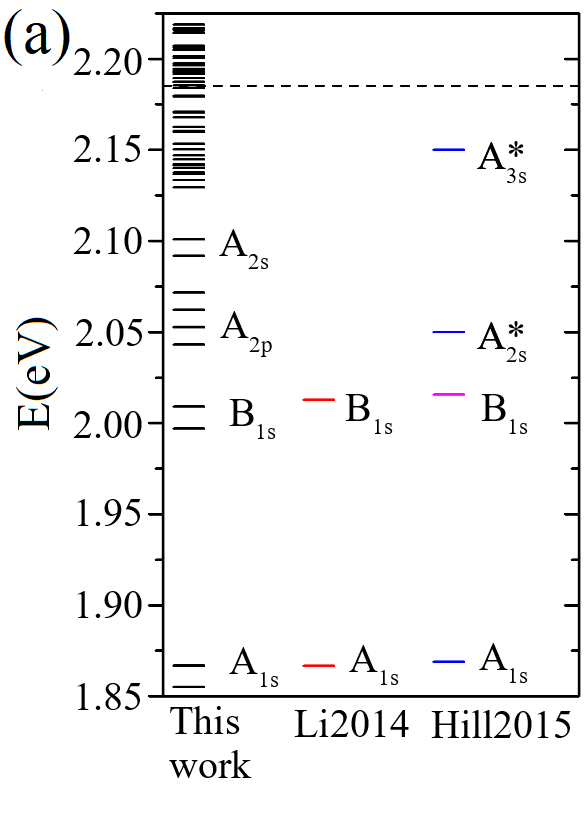}%
\includegraphics[width=0.65\columnwidth]{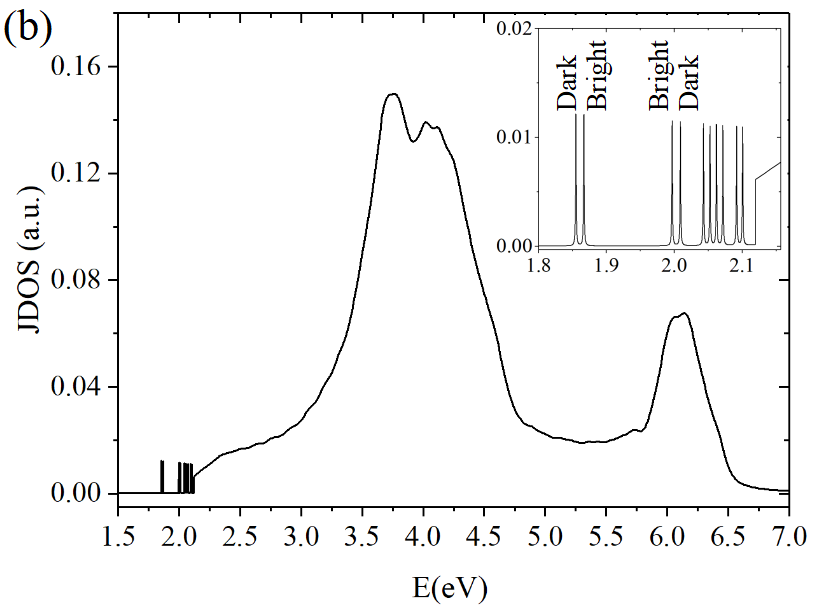}
\caption{
(a) Comparison of the lower portion of the theoretical exciton spectrum 
obtained in this work with the energies of the A and B peaks measured in 
\Refs~\onlinecite{Li2014, Hill2015} (note that the assignment of the peaks 
labeled $\mathrm{A_{2s}^*}$ and $\mathrm{A_{3s}^*}$ is not unequivocal in 
\Ref~\onlinecite{Hill2015}).
After a rigid displacement by $+0.07$\,eV (see text), the theoretical spectrum 
reproduces extremely well the position of the experimental A and B peaks. 
The dashed horizontal line indicates the one-particle energy gap at 2.18\,eV 
($2.18\teq 2.12+0.07$). 
(b) Joint density of exciton states as defined in \Eqref{eq:jdos}, 
already rigidly displaced by $+0.07$\,eV. A Lorentzian level broadening of 
0.1\,eV (full width) has been used, except in the lower energy range (magnified 
in the inset) where it is 0.001\,eV to allow the resolution of individual 
bound exciton levels.
}
\label{fig:spectrum}
\end{figure}

\Fref{fig:spectrum}(a) shows a direct comparison between our BSE-derived 
eigenvalues and experimental spectra measured for MoS$_2$ in silica in the range 
of the single-particle gap \cite{Li2014,Hill2015}. 
We find that, except for a global rigid offset of $0.07$\,eV, the spectrum 
obtained with our model parameters reproduces extremely well the bound exciton 
series. In particular, it captures with accuracy the level spacing of the 
lowest-lying states which are those most sensitive to the modified screened 
potential \eqref{eq:potential} at short distances and, hence, those that most 
clearly deviate from a hydrogen-like spectrum \cite{Srivastava2015, 
Chernikov2014, Maxim2017}. This spectrum also captures the fact that the ground 
excitonic state is dark, as recently established experimentally 
\cite{Molas2017}.
We obtain a bright-dark splitting of $\Delta E_{\rm bd} \approx 12$\,meV for the 
lowest excitonic states. This value is due to the 6\,meV separation of the conduction 
bands due to spin-orbit coupling plus differences in effective masses. 
Our results agree with other theoretical calculations that find 
$\Delta E_{\rm bd} \alt 20$\,meV \cite{Echeverry2016, Baranowski2017, Malic2018} 
depending on the kind of \emph{ab-initio} approach.
So far, the only direct experimental investigation of the bright-dark splitting in MoS$_2$
\cite{Molas2017} finds $\Delta E_{\rm bd} \approx 100$\,meV. This value is unexpectedly
large in view of the theoretical literature and considering that for TMDs with a larger 
spin-orbit coupling, like MoSe$_2$ and WSe$_2$, one finds $ \Delta E_{\rm bd} \approx 
47 - 57$\,meV \cite{Zhang2017, Molas2017}. 
These elements indicate that that quantitative aspects of $\Delta E_{\rm bd}$ in MoS$_2$
are still under experimental scrutiny.

To allow a direct comparison of the absorption spectrum with the experiments, a rigid blue-shift in the 
energies by $+0.07$\,eV has been incorporated in the results shown in 
\Fref{fig:spectrum}. Such a ``calibration'' is somewhat expected because of the 
effective parameterizations of both the band structure and the screened Coulomb 
potential%
\footnote{
Alternatively, the experimental positions of the A and B peaks can be matched 
by tuning the environment dielectric constants that determine $r_0$ [cf. 
\Eqref{eq:r0}] \cite{Pedersen2014}.
}
(yet, a $0.07$\,eV offset is rather small in comparison with similar approaches, 
where  corrections of up to $0.57$\,eV were necessary \cite{Wu2015}). This 
calibration of the energy axis has been also applied in the presentation of all 
our subsequent results.

The bound series is directly associated with transitions between the
topmost valence and bottom conduction bands near the $K$ point of the BZ. Each 
level is 2-fold degenerate on account of the $K{-}K'$ valley degeneracy. When 
the small energy difference between the two lowest spin-polarized conduction 
bands at $K$/$K'$ is ignored, there is an additional 2-fold degeneracy. In our 
calculations, that small energy difference is explicitly finite [cf. 
\Fref{fig:bands}(b)] which explains the existence of two closely spaced levels 
labeled, for example, A$_\mathrm{1s}$, in \Fref{fig:spectrum}(a). Nevertheless, 
one should keep in mind that only half of these excitons are optically bright 
due to the parity selection rule \cite{Wu2015}.

In this respect, it is worth to point out that the difference in the excitation 
energies of the lowest bound A and B excitons does not follow exactly the 
spin-orbit spitting of the bands. This is significant because many theoretical 
TB parameterizations in the literature have identified $E^B \tminus E^A$ 
directly with the SO coupling parameter and, equivalently, this exciton 
splitting is frequently used as a direct experimental measure of the spin-orbit 
splitting in the single-particle band structure, which is not strictly correct. 
For example, we obtain $E^B \tminus E^A \teq 130$\,meV while the valence band 
splitting due to SOC in our chosen TB is 150\,meV. The difference can be mainly 
traced back to the different effective masses of the spin-split valence bands 
\cite{Malic2014}.

To identify the $\bk$-point sampling that guarantees convergence of the 
whole spectrum, we have analyzed the exciton joint density of states (JDOS), 
\begin{equation}
  \rho_J(E) \equiv 
  \frac{1}{N_c N_v N_k^2} \sum_{M} \delta (E-E_M),
  \label{eq:jdos}
\end{equation}
where $E_M$ are the eigenvalues of the BSE. Our calculations show that the JDOS 
is already reasonably converged for $N_k\teq 48$ and that the differences for 
$N_{k}\teq 90$ up to $120$ are negligible. We take $N_{k}\teq 60$ for the 
results presented in this paper. The converged JDOS  calculated with $N_c \teq 
8$ conduction and $N_v \teq 2$ valence bands is shown in \Fref{fig:spectrum}(b) 
for $N_k\teq 60$.

\subsection{Linear optical conductivity}
\label{sec:linear_optical_conductivity}

\begin{figure}[tb]
\centering
\includegraphics[width=\columnwidth]{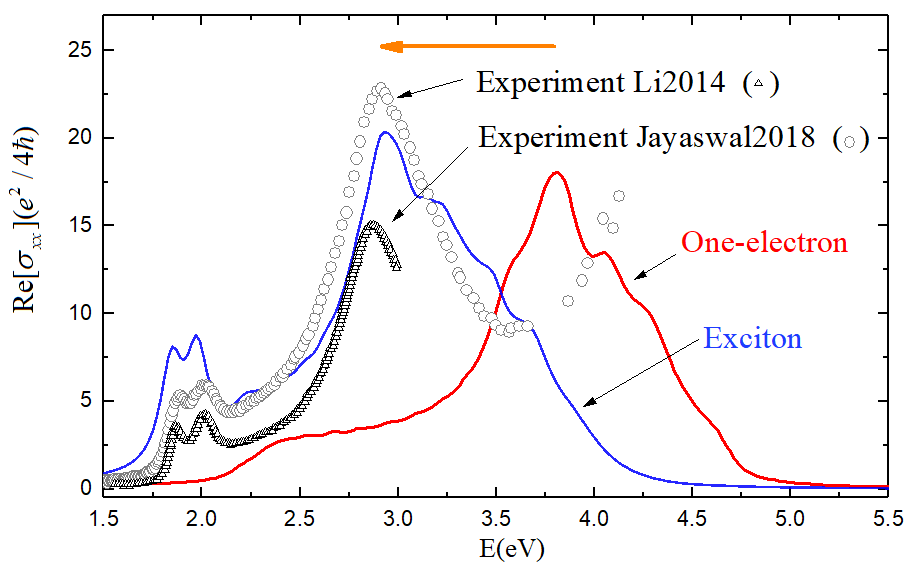}
\caption{
The room-temperature experimental traces reported in references 
\onlinecite{Li2014} and \onlinecite{Jayaswal2018} for the linear optical 
conductivity of a MoS$_{2}$ monolayer on silica (black) and its comparison with 
our results with (blue) and without (red) particle-hole interactions 
($N_{c}\teq 8$, $N_{v}\teq 2$, $N_k\teq 60$). 
An energy-independent Lorentzian broadening of $0.136$\,eV  (full 
width) was added to the calculated $\sigma(\omega)$ to capture the experimental 
broadening at the positions of the A and B excitons.
The decay to zero at high energy is artificial: the restricted number of bands 
used in our diagonalization of the BSE makes the calculated spectra complete 
only up to $\tsim 3.5$\,eV. Nevertheless, for completeness, we show here the 
conductivity in the full range spanned by those bands.
}
\label{fig:sigma}
\end{figure}

The optical conductivity in linear order is directly obtained from 
\Eqsref{eq:sigma-final} and \eqref{eq:sigma-single} by a diagonalization of the 
full BSE matrix in the subspace spanned by the 10 bands mentioned above and a 
suitable sampling of the BZ. The Dirac-delta functions are broadened by 
replacing them with Lorenzians of full width $\gamma$. The latter qualitatively 
represents a total decay rate due to several microscopic mechanisms, each of 
them contributing to $\gamma$ with a characteristic energy dependence 
\cite{Qiu2013}. Since the experimental broadening is largely disorder and sample 
dependent, we take the simple approach of considering $\gamma$ as constant and 
to adjust its value to fit the experimental data (see discussion below).

Figure~\ref{fig:sigma} shows the real part of $\sigma(\omega)$ that we obtain 
for the single-particle and BSE calculations. It is our most significant result.
For reference, the plot includes the experimental traces reported recently by Li 
\emph{et al.} \cite{Li2014} for exfoliated MoS$_2$ and Jayaswal \emph{et al.}  
\cite{Jayaswal2018} for CVD-grown MoS$_2$, both measured on silica at room 
temperature. Note that the comparison with experimental data is only meaningful 
up to $\tsim 3.5$\,eV, beyond which excitations involving additional valence 
bands not included in our present calculation (\Fref{fig:bands}a) must be taken 
into account \cite{Ridolfi2015}. We put $\gamma \teq  0.068$\,eV in our 
calculations, which corresponds to the average width of the A and B exciton 
peaks reported in the experiment of Li \emph{et al.}.
As anticipated, the single-particle results fail to describe the spectral 
features of the optical response. Even though the discrepancy is most obvious in 
the region dominated by the bound exciton levels, $E \lesssim 2.1$\,eV, there is 
also a remarkable difference in the continuum. In particular, the 
single-particle spectral weight is generically distributed at energies much 
above the interaction-corrected values, as highlighted by the horizontal arrow 
in Figure~\ref{fig:sigma}. Therefore, by neglecting the excitonic effects and 
interpreting the single-particle absorption spectrum literally, one incurs in a 
severe qualitative and quantitative misrepresentation, not only in the vicinity 
of the optical gap, but actually over the entire range of energies. Thus, the 
strong Coulomb interactions in these two-dimensional materials, not only lead to 
high exciton binding energies, but also largely renormalize the whole spectrum.

Figure \ref{fig:sigma} shows that our BSE calculation captures rather well the 
important experimental features of the optical conductivity in MoS$_2$, namely, 
the position of the A and B peaks, the overall maximum at $E^C \tsimeq 
2.85$\,eV, the energy dependence over the entire experimental range, and the 
absolute magnitude of the optical conductivity. We stress that we do not adjust 
or scale the magnitude of $\sigma(\omega)$, as is frequently done in 
theoretical work that discusses similar comparisons with experimental data.  
This overall agreement attests to the validity and accuracy of the SK 
parameterization of the underlying band structure, and provides strong 
support to the use of the Keldysh effective screened potential in the BSE 
calculations (an additional overview of different theoretical and experimental 
spectra reported in recent literature is given in \Fref{fig:sigma-comparisons} 
of the appendix).
 
Since $\gamma$ corresponds to microscopic processes that depend on 
the excitation energy, one might expect such dependence to have significant 
influence in the line shape of the optical conductivity. Yet, our calculation 
that uses a constant broadening captures the measured energy dependence quite 
satisfactorily. Surprisingly, \emph{ab initio} results \cite{Qiu2013, 
Ramasubramanian2012, Klots2014} are less accurate in describing 
$\Re\sigma(\omega)$ despite the inclusion of specific energy dependent 
broadening processes, such as electron-phonon scattering \cite{Qiu2013} [see 
\Fref{fig:sigma-comparisons}(b)]. This indicates that the experimental 
broadening is likely dominated by disorder and justifies \emph{a posteriori} 
our choice of an energy-independent $\gamma$ to broaden the numerically 
discrete spectrum in \Eqref{eq:sigma-final}.

Finally, as pointed out in \Ref~\onlinecite{Qiu2013}, convergence studies up to 
large $\bk$-sampling meshes are essential to guarantee that one meaningfully 
``\emph{reproduces features in the experimental absorption spectrum above 
2\,eV}''. The comparison presented in \Fref{fig:sigma-comparisons} provides 
a good example of how approaches based on parameterized SK models such as ours 
can outperform full first-principles solutions of the BSE when it comes to 
expediency and the need of a large BZ sampling to ensure convergence. It is key, 
of course, to rely on underlying band structures that accurately describe the 
quasiparticle renormalization from the outset.

\subsection{Nature of the C excitons}

\begin{figure}[tb]
\includegraphics[width=0.65\columnwidth]{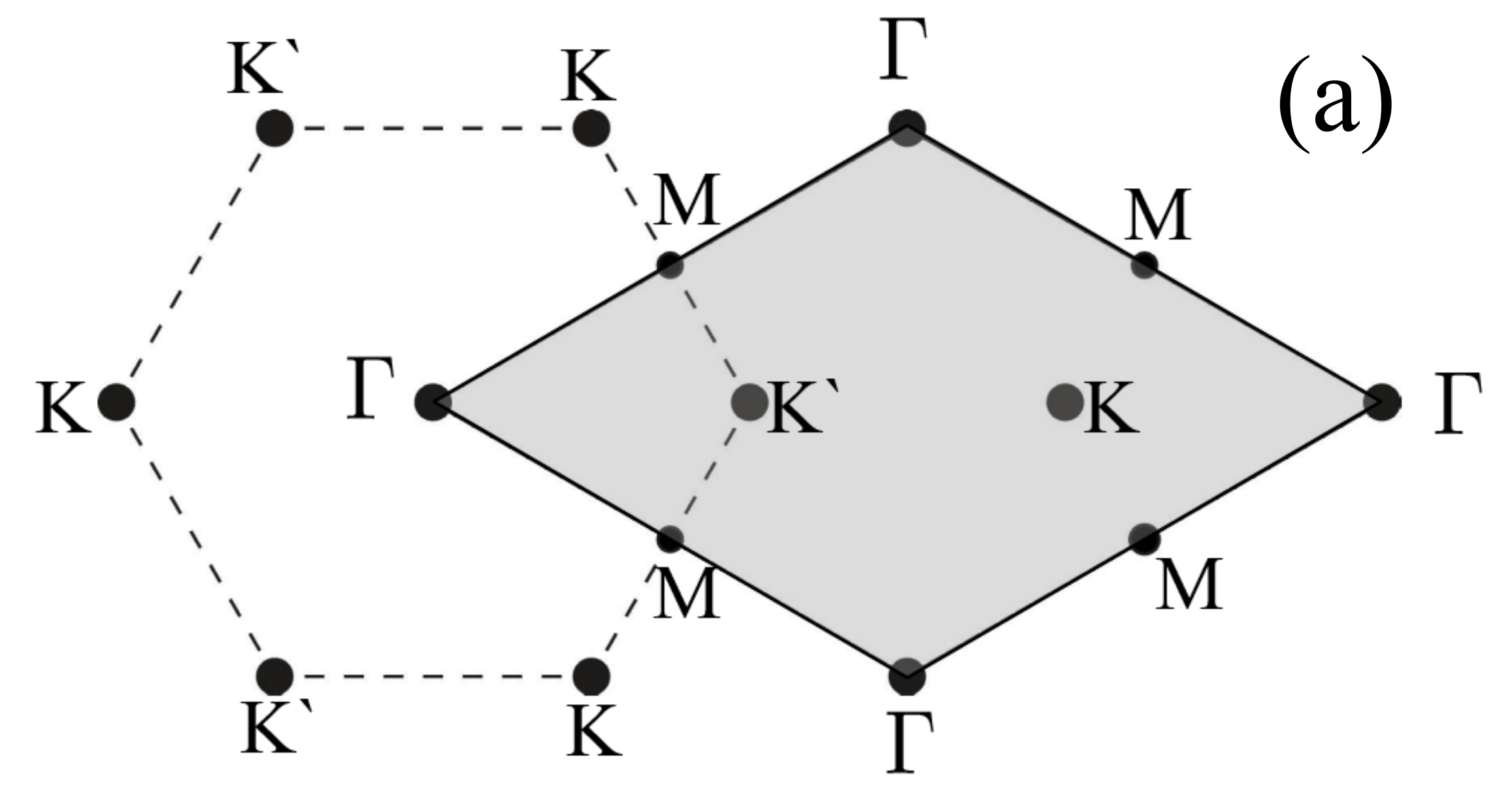} 
\par\medskip
\includegraphics[width=\columnwidth]{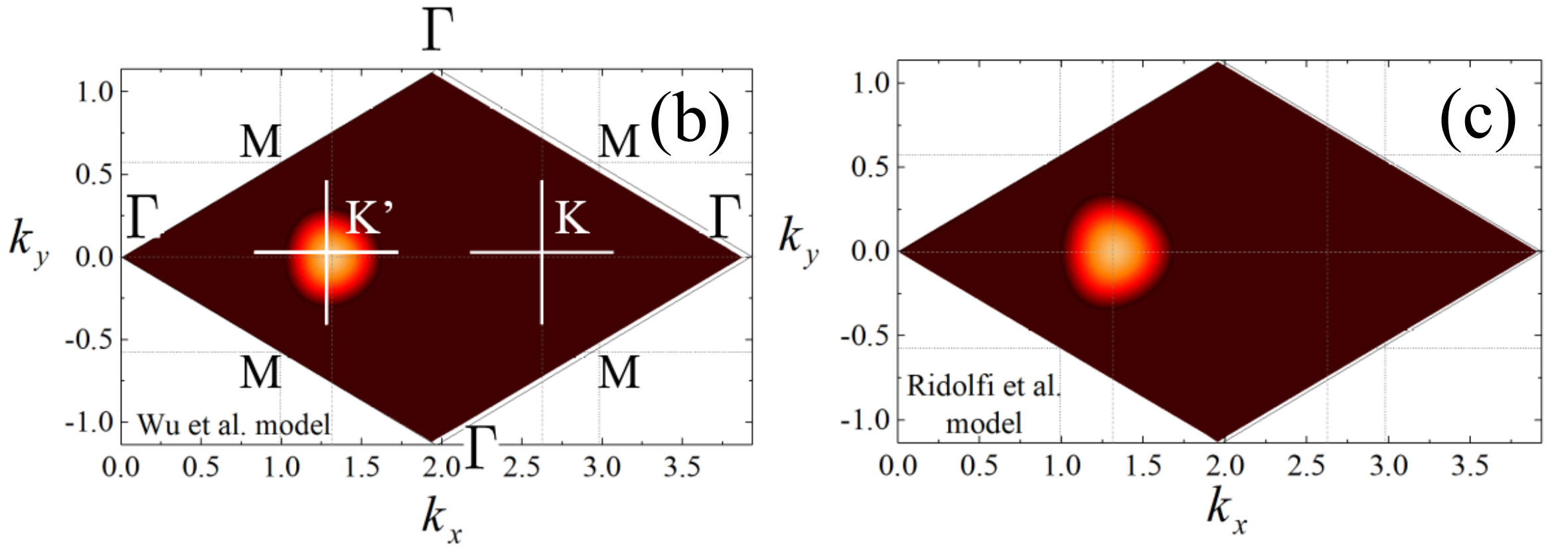} 
\par
\includegraphics[width=\columnwidth]{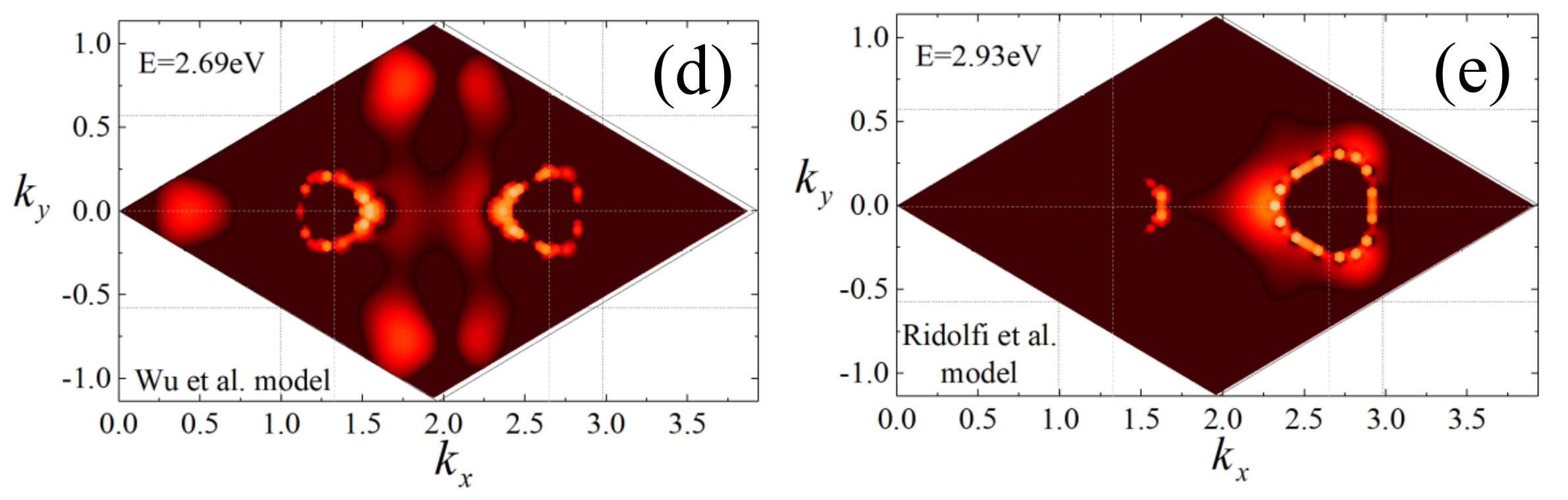}
\par
\includegraphics[width=\columnwidth]{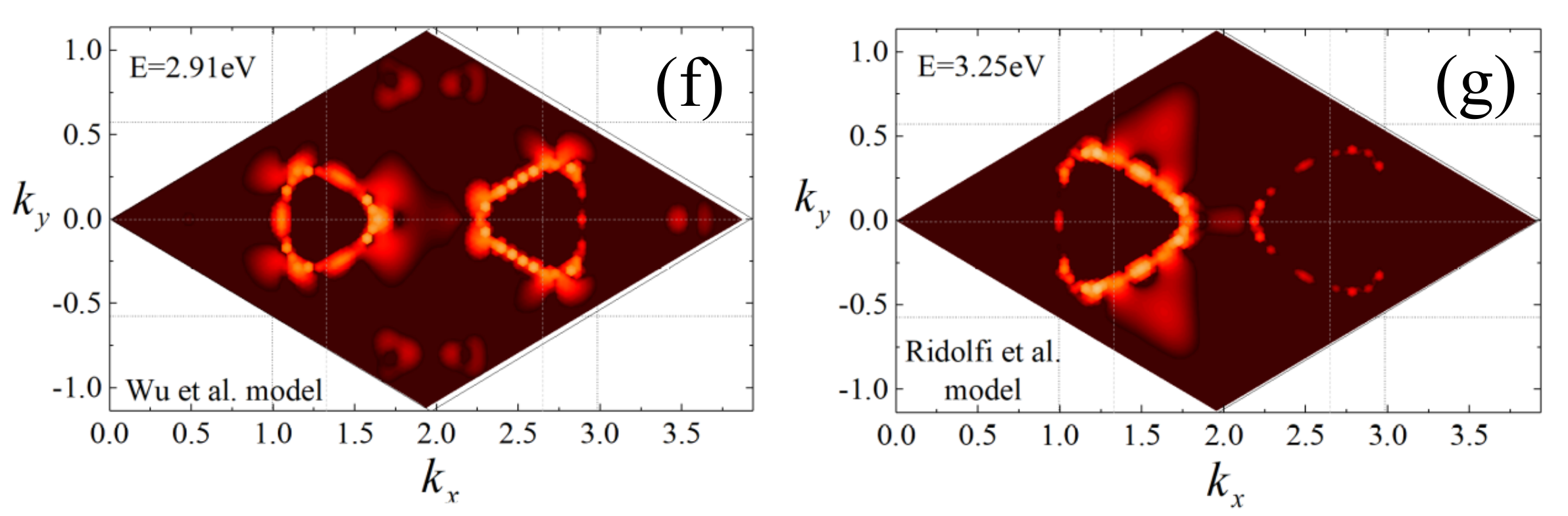}
\par
\includegraphics[width=0.8\columnwidth]{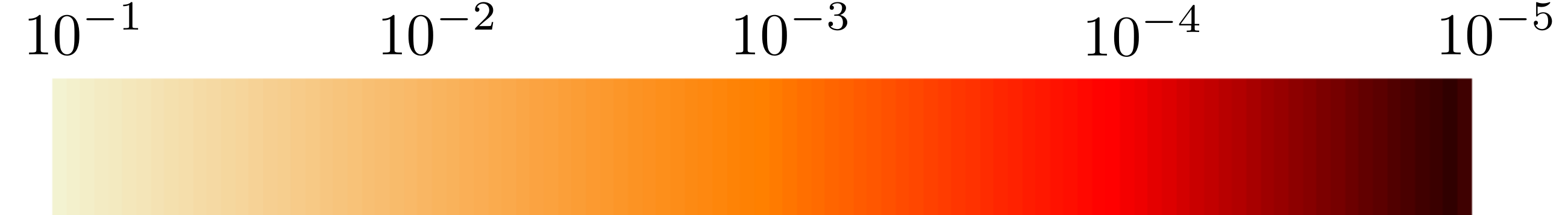}
\caption{
Representative excitonic wave functions in momentum space including only the 
two conduction and valence bands closest to $E_F$ ($N_k\teq 60$, $N_c\teq 
N_v\teq 2$). (a) Diagram of the first BZ of MoS$_2$ (dashed line) and the 
equivalent reciprocal unit cell used in our $\bk$-sampling (solid rhombus). The 
$\Gamma$ point is located at the vertices of the sampling domain, $M\teq 
(\pi/a,-\pi/\sqrt{3}a)$, $K'\teq (4\pi/3a, 0)$ and $K\teq (8\pi/3a, 0)$. 
The left column gives wave functions obtained with the TB model of 
Ref.~\onlinecite{Wu2015}, while those on the right use the SK parameterization 
of Ref.~\onlinecite{Ridolfi2015}. The second row [(b) and (c)] shows density 
plots of the probability distribution arising from one of the wave functions 
associated with the A peak in the optical  conductivity. The two bottom rows 
[(d) to (g)] show representative wave functions 
in the region of C excitons. 
The axes are in units of \AA$^{-1}$. Note that the color scale is logarithmic.}
\label{fig:wavefunctions}
\end{figure}

The maximum in $\Re\sigma(\omega)$ at $E^C$ has been attributed to resonant 
excitons involving transitions near the center of the BZ (the $\Gamma$ point), 
the so-called C excitons \cite{Qiu2013}. As we discuss below, these excitons are 
actually not more related to $\Gamma$ than they are to the $K$ point. Hence, it 
is incorrect to refer to them as ``$\Gamma$-point excitons''.
Turning our attention to the spectral details of the conductivity around $E^C$,  
\Fref{fig:sigma} shows that the energy dependence obtained 
from the excitonic calculation in the interval $[2.5,\,3.5]$\,eV is nearly 
identical to that given at the single particle level in the different interval 
$[3.5,\,4.5]$\,eV. This seems to indicate that, in this energy range, the 
primary effect of the electronic interaction is to rigidly redshift the 
one-electron conductivity by about $0.9$\,eV, without notable modifications to 
the line shape (indicated by the horizontal arrow in \Fref{fig:sigma}).
That being the case, one could question the attribution of the enhanced 
spectral weight in this broad region to interaction effects, insofar as 
(i) the one-electron trace seems to already carry the key aspects of the energy 
dependence and magnitude of $\sigma(\omega)$ and 
(ii) the excitonic corrections do not seem to generate any additional spectral 
feature beyond simply repositioning the curve \emph{en bloc} to lower energies, 
as expected from the attractive electron-hole interaction. 
In other words, it appears as if, for excitation 
energies belonging to the one-particle continuum, the restructuring of the 
absorption spectrum caused by interactions amounts to a ``scissor''-type 
correction of the energy spectrum, where an effective ``binding energy'' of 
$\tsim 0.9$\,eV brings the one-electron trace (red) to its correct position 
(blue), but with essentially no changes in oscillator strength. The 
problem with this, however, is that the value $0.9$\,eV is much larger than the 
binding energy of the A \emph{bound} excitons ($E^A_b\teq 0.324$\,eV). It 
turns out that explaining this excitonic redshift on the basis of these 
one-electron band structure features is questionable and, as we now explain in 
detail, too simplistic and misleading. 

The large spectral weight of the one-electron curve (red in \Fref{fig:sigma}) 
around its maximum is primarily due to the downward dispersion of the lowest 
conduction bands along $\Gamma{-}K$ (see \Fref{fig:bands}): the fact that 
conduction and valence bands separated by excitation energies $\tsim 4$\,eV 
disperse roughly parallel to each other entails a large enhancement of the 
one-particle JDOS in this region, naturally explaining the peak in $\Re 
\sigma(\omega)$. Equivalently, it can be inferred from \Fref{fig:bands} that, 
in our SK model, the one-electron ``optical band structure'' is nearly flat 
along the $\Gamma{-}K$ line. Hence, the one-electron ``C peak'' is mostly the 
result of a large one-electron JDOS at $\tsim 4$\,eV (see also 
\Fref{fig:sigma-jdos-small-broadening} in the appendix which explicitly 
confirms this). 

However, the same conclusion does not apply to the results of the excitonic 
calculation. As we have seen in \Fref{fig:spectrum}(b), the excitonic JDOS is 
peaked, broadly speaking, at $\tsim 4$\,eV while the corresponding conductivity 
peaks at $E^C \tsimeq 2.85$\,eV. It follows that the enhanced optical response 
near $E^C$ is, clearly, not the result of a large number of excitonic levels 
with energies close to $E^C$. In reality, except for the bound excitonic levels 
that emerge in the gap, the one-particle and excitonic JDOS do not differ much 
at energies in the continuum, as we demonstrate in 
\Fref{fig:sigma-jdos-small-broadening} (appendix). For example, predicting the 
spectral shape of the optical conductivity solely on the basis of the excitonic 
spectrum (through the JDOS) would clearly fail for the energies in the 
continuum. This is the reason why the idea of an effective ``binding energy'' of 
$\tsim 0.9$\,eV that rigidly redshifts the one-electron spectrum, as discussed 
above, is misleading.

We must therefore explicitly consider the oscillator strengths which, 
according to \Eqref{eq:sigma-final}, are given by 
\begin{equation}
  \biggl|
    \sum_{\bk cv}A_{cv\bk}^M \sum_{\alpha \beta} 
      (C_{\alpha \bk}^v)^* C_{\beta \bk}^c 
      \nabla_{k_x} \! \bra{\phi_\alpha} \hat{H}(\bk) \ket{\phi_\beta}
  \biggr|^2.
  \label{eq:osc-str}
\end{equation}
Recalling that $A_{cv\bk}^M$ represents the probability amplitude of the 
exciton $M$ in reciprocal space, the oscillator strength depends not only on the 
one-electron dipole matrix elements, but also on the specific texture of each 
excitonic wave function in $\bk$ space. 

In \Fref{fig:wavefunctions} we analyze representative excitonic wave functions in 
reciprocal space associated with the largest optical spectral weights (cf.  
\Fref{fig:sigma} and see also \Fref{fig:osc-str} below). As per our earlier 
remarks regarding the ``calibration'' of the energies, the values indicated in 
each panel are shifted by $+0.07$\,eV (right column) and $+0.57$\,eV (left 
column) with respect to the original eigenvalues of the BSE.
For reference, \Fref{fig:wavefunctions}(b) and \ref{fig:wavefunctions}(c) show 
that the wave functions associated with the A and B excitons  concentrate at 
the vicinity of the $K$ points
\footnote{The results for the B excitons are not shown explicitly, but are 
similar to those reported in \Fref{fig:wavefunctions} for the A counterparts.},
as has been well established by previous calculations. The degree of their 
localization is directly related to the large binding energies and, in real 
space, they appear much more localized than typical excitons in semiconductors 
\cite{Wu2015, WangReview2017}.
We recall that, due to the parity selection rule, only half of the excitons 
related to the two valence and two conduction bands straddling the gap are 
optically bright. Each of the optically bright excitons is 2-fold degenerate 
on account of the $K$--$K'$ valley degeneracy (when the small energy difference 
between the two lowest spin-polarized conduction bands at $K$/$K'$ is ignored, 
these are further degenerate with the other pair of doubly-degenerate dark 
excitons. In our calculations, that small energy difference is explicitly 
finite, in correspondence with the spectrum shown in 
\Fref{fig:spectrum}).

The picture is rather different for the C excitons. The wave functions shown in 
panels (d) to (g) of \Fref{fig:wavefunctions} correspond to selected states with 
energies close to $E^C$. We caution the reader that, unlike the case of bound 
excitons such as A or B, here we show selected representative wave functions of 
a ``continuum'' of states. (We have checked over a window of energies near $E^C$ 
that the spreading of the wave functions over portions of the BZ similar to 
those shown here is a robust common feature.) Obviously, the two TB models we 
use \cite{Wu2015,Ridolfi2015} render different states. We note, however, the 
corresponding wave functions show rather large similarities 
(compare the left and right columns in \Fref{fig:wavefunctions}).
Remarkably, there is no pronounced contribution \emph{right at} the $\Gamma$ 
point itself; on the contrary, for our choice of primitive cell, the excitonic 
wave functions appear distributed on a ring at a finite distance from all the
symmetry points, midway between $K{-}M$ and $\Gamma{-}K$. This agrees 
with similar observations based on full \emph{ab initio} solutions of the BSE in 
the region of the C excitons which, likewise, associate C excitons with 
transitions within a similar annular region, but not specifically at or near 
$\Gamma$ \cite{Qiu2013,Klots2014,Molina-Sanchez2015} (see, for example, Fig. 21 
in \Ref~\onlinecite{Molina-Sanchez2015}). Further, the C excitons in 
\Fref{fig:wavefunctions} appear as more tightly localized in $\bk$ space than 
their A counterparts, since the rings are rather thin (keep also in mind that 
the color scale is logarithmic). 

\begin{figure}[tb]
\includegraphics[width=\columnwidth]{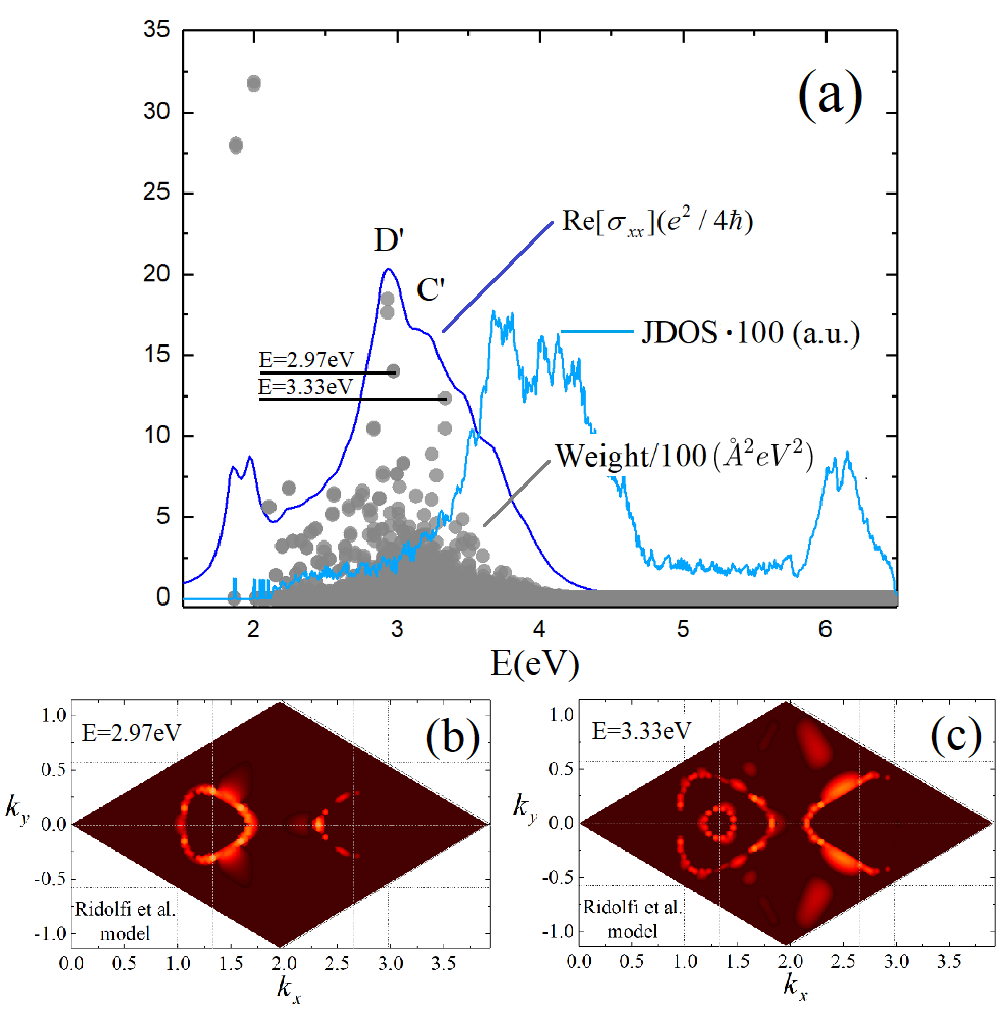} 
\caption{
(a) An overlay of the real part of the optical conductivity, the excitonic JDOS 
and the oscillator strength expressed in \Eqref{eq:osc-str} according to our 
excitonic calculation. 
(b) and (c) show representative excitonic wave functions contributing to the 
peaks D' and C' whose energies and spectral weight are highlighted in panel 
(a). ($N_k\teq 60$, $N_c\teq 8$, $N_v\teq 2$). 
The axes are in units of \AA$^{-1}$. Note that the color scale is logarithmic 
and the same as in \Fref{fig:wavefunctions}. 
}
\label{fig:osc-str}
\end{figure}

This localization is established at a quantitative level by a computation of 
the inverse participation ratio associated with each exciton wave function, 
which we describe in the appendix. The data shown there, in 
\Fref{fig:sigma-jdos-small-broadening}(b), reveal that while
the region of the C excitons is characterized by a comparatively small JDOS, 
the corresponding states are typically more localized than the average
\footnote{
At a technical level, it is important to realize that the sharp localization 
associated with the C excitons requires a fine $\bk$-point mesh to ensure 
proper convergence of the absorption spectrum over this large range of energies 
\onlinecite{Qiu2013}.
}.
This ultimately determines the energy dependence of the oscillator strength, 
which is maximized in the region of energy near $E^C$, as can be directly seen 
in \Fref{fig:osc-str} where the quantity \eqref{eq:osc-str} is shown for all 
excitons. Therefore, opposite to the case of a single-particle calculation, the 
spectral profile of $\Re\sigma(\omega)$ is almost entirely determined by the 
oscillator strength and not the optical JDOS.

In conclusion, the discussion above indicates that it is misleading to designate 
these as ``$\Gamma$-point excitons'' and reinforces the perspective that 
relates them with the properties of the ``optical band structure'' along the 
$\Gamma{-}K$ and $K{-}M$ directions \cite{Qiu2013, Klots2014}. 
Extending the calculation of the excitonic wavefunctions shown in 
\Fref{fig:wavefunctions} to include not only the lowest 2 but all 8 conduction 
bands in our TB model, we can conclusively assign the two peaks in the 
oscillator strength at $E \tsimeq 2.9$\,eV and $E \tsimeq 3.3$\,eV (labeled as 
D' and C' in \Fref{fig:osc-str}) to the two contributions distinguished in 
\Ref~\onlinecite{Aleit2016}.
Specifically, with our band structure, they arise from particle-hole excitations 
between approximately parallel bands $\tsimeq 3.8$\,eV apart along the 
$\Gamma{-}K$ and $K{-}M$ symmetry lines. From this point of view, excitons 
belonging to the broad C region do have a large binding energy of about 
$0.9$\,eV ($0.9 \teq 3.8 \tminus 2.9$). 
Notably, a comparison between \Fref{fig:wavefunctions}(g) and 
\Fref{fig:osc-str}(c) reveals additional weight in the latter over an inner ring 
close to $K$. This is contributed by transitions from the valence to the 5th 
and 6th conduction bands which disperse downwards and nearly parallel to each 
other $\tsimeq 4$\,eV apart near $K$. This vividly illustrates that the 
attribution of fine details associated with the whole region of the C excitons 
is sensitive to the particulars of the underlying bandstructure, and 
necessitates the inclusion of higher conduction bands.

That C excitons arise from particle-hole excitations midway from the 
$\Gamma{-}K$ and $K{-}M$ lines in reciprocal space is consistent with the C 
peak being more sensitive to the number of layers in thin MoS$_2$ films than the 
A and B features. In particular, the experimental shift of $E^C$ correlates with 
the changes in the separation of bands with MoS$_2$ thickness 
\cite{Malard2013,Kim2014,Klots2014}. These changes in electronic structure are 
known to be small at the $K$ point but large along the whole $\Gamma{-}K$ line, 
ultimately determining the transition from a direct to indirect gap as a 
function of thickness \cite{Splendiani2010}. The contrast between A/B and C 
excitons can be understood from the fact that the electronic states at $K$/$K'$ 
contain mostly contributions from the $d$ orbitals in Mo, while in the regions 
of $\bk$ that contribute to the C excitons they have a strong $p$ character 
arising from the sulfur atoms 
\cite{Cappelluti2013,Molina-Sanchez2015,Ridolfi2015}.
As $d$ orbitals are spatially more localized and, moreover, lie in the inner of 
the 3 atomic planes that make each MoS$_2$ monolayer, the A and B exciton states 
at $K$/$K'$ are not as perturbed in a stacked multilayer structure or as a 
result of strain, in comparison with the changes that occur to the C excitons 
due to their strong sulfur orbital content \cite{Molina-Sanchez2015}.

\section{Summary}
\label{sec:conclusions}

We revisited the problem of calculating the excitonic spectrum in the MoS$_2$ 
monolayer. It has been shown that many-body effects strongly restructure 
the optical absorption spectrum over an unusually large range of energies in 
comparison with the single-particle picture.  
Our approach accounts for the anomalous screening in two dimensions and for the 
presence of a substrate, both modeled by a suitable effective Keldysh 
potential. We solve the Bethe-Salpeter equation for the interacting 
electron-hole excitations by using a Slater-Koster tight-binding model 
parametrized to fit the calculated first-principles band structure of the 
material. The optical conductivity that emerges captures with good accuracy both 
the shape and absolute magnitude of the experimental data. 

Our calculation does not consider any temperature-induced change in the band 
structure nor microscopic broadening mechanisms such as those from the 
unavoidable phonon excitations. Indeed, by solving a temperature-dependent BSE 
based on first-principles electron and phonon spectra, Molina-Sanch\'ez \emph{et 
al.} \cite{Molina2016} have shown that the electron-phonon coupling is 
responsible for most of the 40\,meV red-shift observed between zero and room 
temperature \cite{Kioseoglou2016}. In addition to this, for quantitative and 
qualitative accuracy at the microscopic level, one must consider the impact of 
the thermal expansion in the band structures and, in the case of excitons, the 
broadening contributed by radiative recombination. Details of such processes 
are, however, not in the scope of the present work; in many experimental cases, 
such microscopic details are overwhelmed by disorder-induced broadening. To 
compare our results with experiments at room temperature, we blue-shifted the 
calculated optical response spectrum by 70\,meV and introduced a 
phenomenological broadening, as discussed in Sec.~\ref{sec:results}.

Seeing that our result captures well the experimental spectrum up to $\tsim 
3.5$\,eV, we relied on the predictions of this model to investigate the effects 
and characteristics of the so-called C excitons. Notably, we explicitly showed 
in \Fref{fig:wavefunctions} that they arise from particle-hole excitations in 
an annular region of the BZ centered at, but at finite radii from, the $K$ 
points (maximal contributions arise from regions between $\Gamma{-}K$ and 
$M{-}K$).

The interplay between the texture of the excitonic wave functions and the 
one-electron dipole matrix elements is responsible for the massive transfer of 
spectral weight seen in the conductivity when compared with results at the 
one-electron level (\Fref{fig:sigma}). Our results also suggest a cautionary 
word when it comes to effective mass descriptions of the MoS$_2$ band structure, 
especially if the aim is to describe the optical excitations in the vicinity of 
$E^C$. In this case, a model that captures only the band structure at the 
$\Gamma$ point will be certainly insufficient for that. 

Throughout our analysis, we presented results obtained using two different 
tight-binding descriptions of the underlying single-particle band structure. 
This provides an example of the immediate transferability in this approach to 
study the optical response of other members of the TMD family. In such a case, 
one can readily use the same orbital basis for the TB model and the only 
material-dependent input is the corresponding band structure. The generic 
workflow is the same as the one we have used above: (i) Obtain an 
\emph{accurate} quasiparticle band structure from first principles, (ii) 
Determine the parameters of the Slater-Koster TB that most faithfully 
describe such band structure, and (iii) Solve the BSE in the eigenbasis of that 
TB Hamiltonian. 

Our analysis of two different TB parameterizations also affords a perspective 
over some aspects that are robust in this approach and others that depend on 
fine details of the parameterization. Overall, the accuracy of our results based 
on the SK model developed in \Ref~\onlinecite{Ridolfi2015} vividly supports the 
use of effective models to expeditiously explore the properties of excitons in 
2D materials. This work shows it to be a reliable strategy, provided that the 
starting Hamiltonian faithfully describes the quasiparticle-corrected band 
structure. These approaches are orders of magnitude faster in CPU time than 
complete first-principles solutions of the BSE. Such an advantage facilitates 
properly addressing the optical response of MoS$_2$ at energies around the C 
excitons, where a fine sampling of the $k$-space is necessary. We believe that, 
due to their intrinsic flexibility to model reliably a variety of conditions 
such as heterostructures, disorder and strain, effective models open the path 
for a more comprehensive investigation of the optical properties of TMDs where 
interaction effects play a fundamental role.

\begin{acknowledgments}
We acknowledge fruitful discussions with P. E.~Trevisanutto, V. Olevano, 
T.~G. Pedersen, L. Lima, F.~Wu, F.~Qu, and M.~L. Trolle.
E. Ridolfi was supported by the Singapore Ministry of Education under grant 
number MOE2015-T2-2-059.
This work was further supported by the Singapore Ministry of Education Academic 
Research Fund Tier 1 under Grant No. R-144-000-386-114 and the Brazilian funding agencies CNPq, CAPES, and FAPERJ.
Numerical computations were carried out at the HPC facilities of the NUS Centre 
for Advanced 2D Materials.
\end{acknowledgments}

\section*{Appendix}

\subsection{Optical conductivity: compilation of theoretical results}

\begin{figure}[!htb]
\centering
\includegraphics[width=0.95\columnwidth]{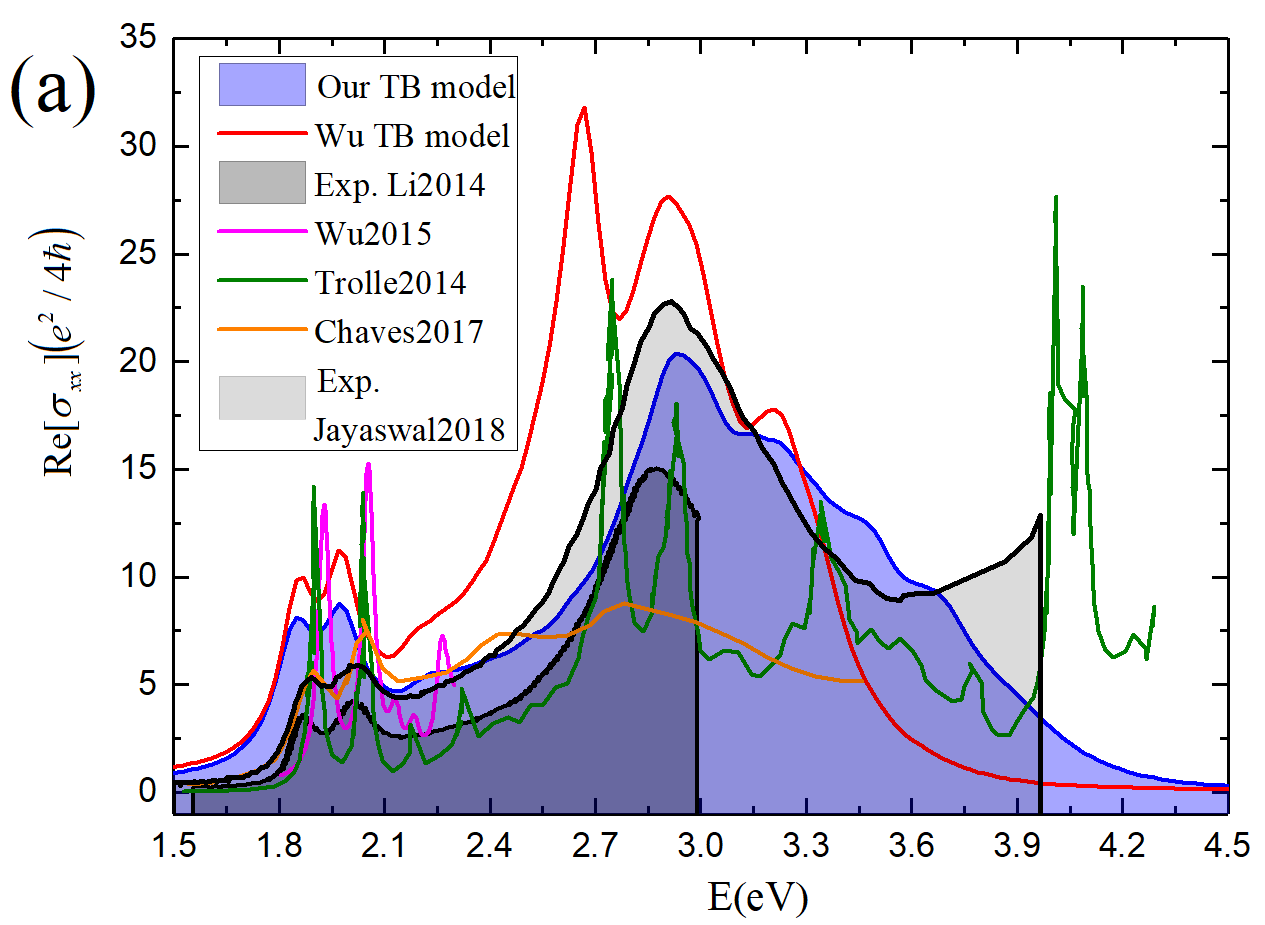} \\
\includegraphics[width=0.95\columnwidth]{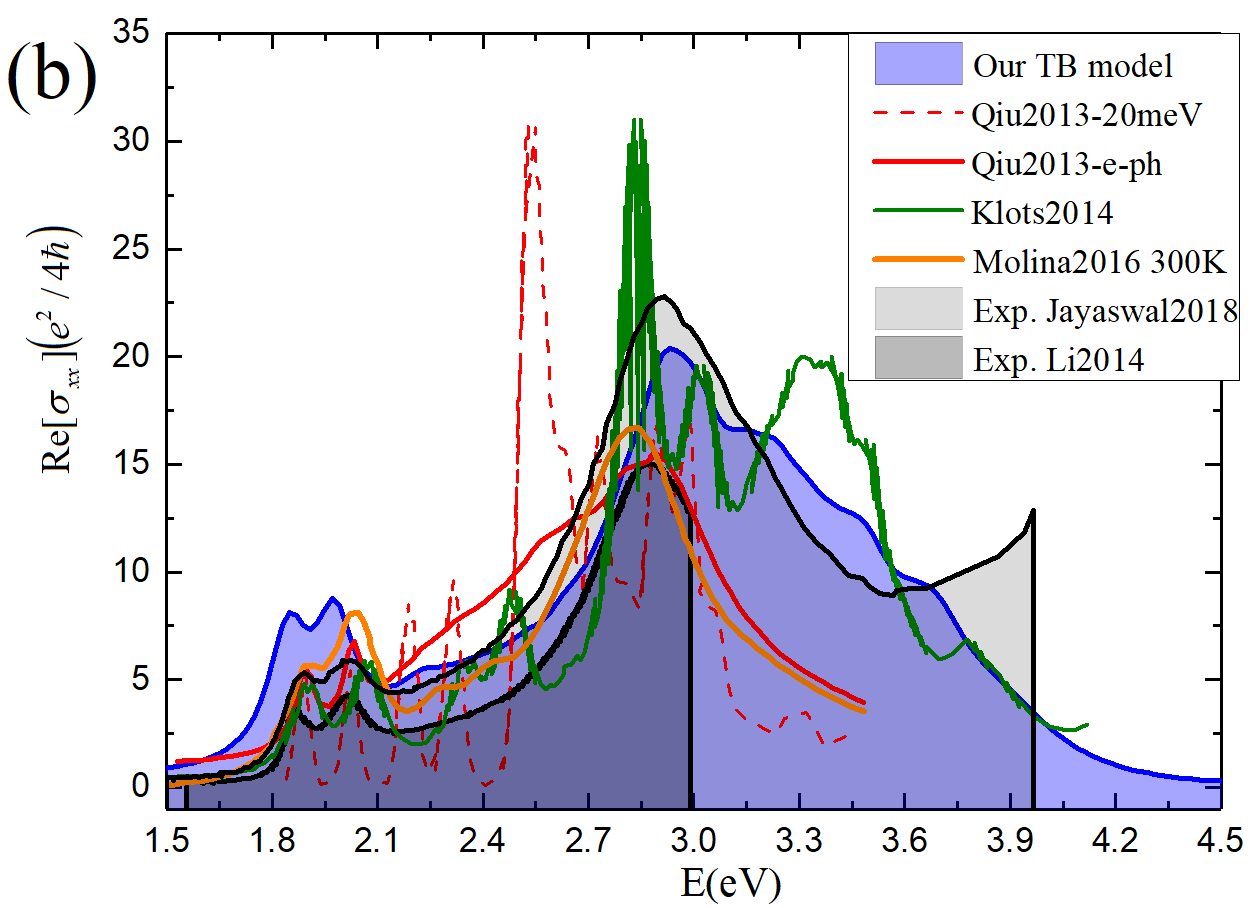}
\caption{
Comparison between different theoretical predictions for the optical 
conductivity and the experimental results of Li and collaborators \cite{Li2014} and Jayaswal and collaborators \cite{Jayaswal2018}. 
(a) Re($\sigma_{xx}$) obtained from the TB-based solutions of the BSE reported 
in \Refs~\onlinecite{Wu2015,Pedersen2014,Nuno2017} contrasted with our  
calculations based on both the TB model of Ridolfi \emph{et al.} 
\cite{Ridolfi2015} (blue) and that of Wu \emph{et al.} \cite{Wu2015} (red) (in 
the latter we added the same broadening indicated in \Fref{fig:sigma}; this 
curve must be rigidly shifted by $+0.57$\,eV to match the experimental position 
of the A and B excitons \cite{Wu2015}). 
(b) Re($\sigma_{xx}$) based on full \emph{ab initio} solutions of the BSE from 
\Refs~\onlinecite{Qiu2013, Klots2014, Molina2016}. Since the latter 
\cite{Klots2014, Molina2016} present Re($\sigma_{xx}$) in arbitrary units, we 
vertically scaled each curve to directly compare with the experimental trace.
The curve from \Ref~\onlinecite{Molina2016} is at 300\,K.
}
\label{fig:sigma-comparisons}
\end{figure}

Figure \ref{fig:sigma-comparisons} gives an overview of different theoretical 
results for the optical conductivity in MoS$_2$ found in the literature. Panel 
(a) compiles TB-based calculations \cite{Pedersen2014, Nuno2017, Wu2015}, while 
panel (b) compares \emph{ab initio} results 
\cite{Molina2016,Qiu2013,Klots2014}. 

In the case of the TB model, \Fref{fig:sigma-comparisons} shows both
the $\Re\sigma(\omega)$ given in \Ref~\onlinecite{Wu2015} as well as 
the our BSE calculation using Wu and collaborators 
 \cite{Wu2015}  TB parameterization over a wider energy range with a suitable
broadening adapted to the experimental traces. 
Figure \ref{fig:sigma-comparisons} indicates that the different theoretical 
approaches describe roughly the same behavior of $\Re\sigma(\omega)$ around the 
$A$ and $B$ peaks%
\footnote{
We stress, however, that some models require rather large rigid shifts in energy 
and/or vertical scaling in order to make the calculated results agree with the 
experimental traces as shown in \Fref{fig:sigma-comparisons}. 
}.
In contrast, the energy dependence and spectral weight in the interval 
$[2.0,\,3.0]$\,eV that covers the region of the $C$ excitons are significantly
approach dependent. 
In the particular case of the two TB models that we analyze in detail, these 
differences can be traced to the larger splitting of the conduction bands near 
the $\Gamma$ point in the model of \Ref~\onlinecite{Wu2015} and the different 
orbital content of the Bloch states that dominate the dipole matrix 
elements
\footnote{
For completeness, it is worth noting that the spectral shape in the range of the 
$C$ excitons measured in MoS$_2$ multilayers changes appreciably with layer 
number, as shown by the experiments reported in \Ref~\onlinecite{Kim2014}.
}. 

\subsection{Number of conduction bands}

\begin{figure}[htb]
\centering
\includegraphics[width=0.9\columnwidth]{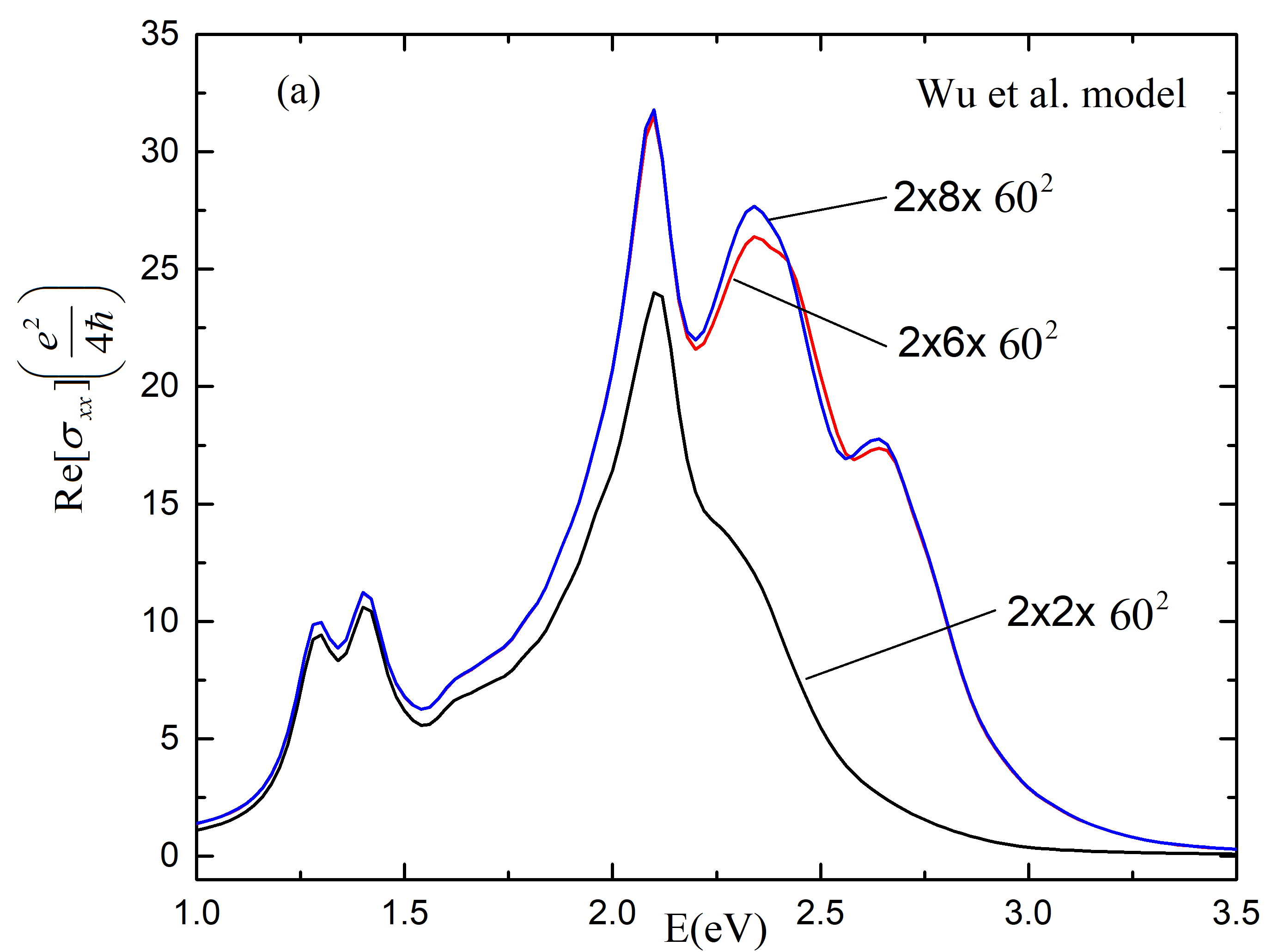} \\
\includegraphics[width=0.9\columnwidth]{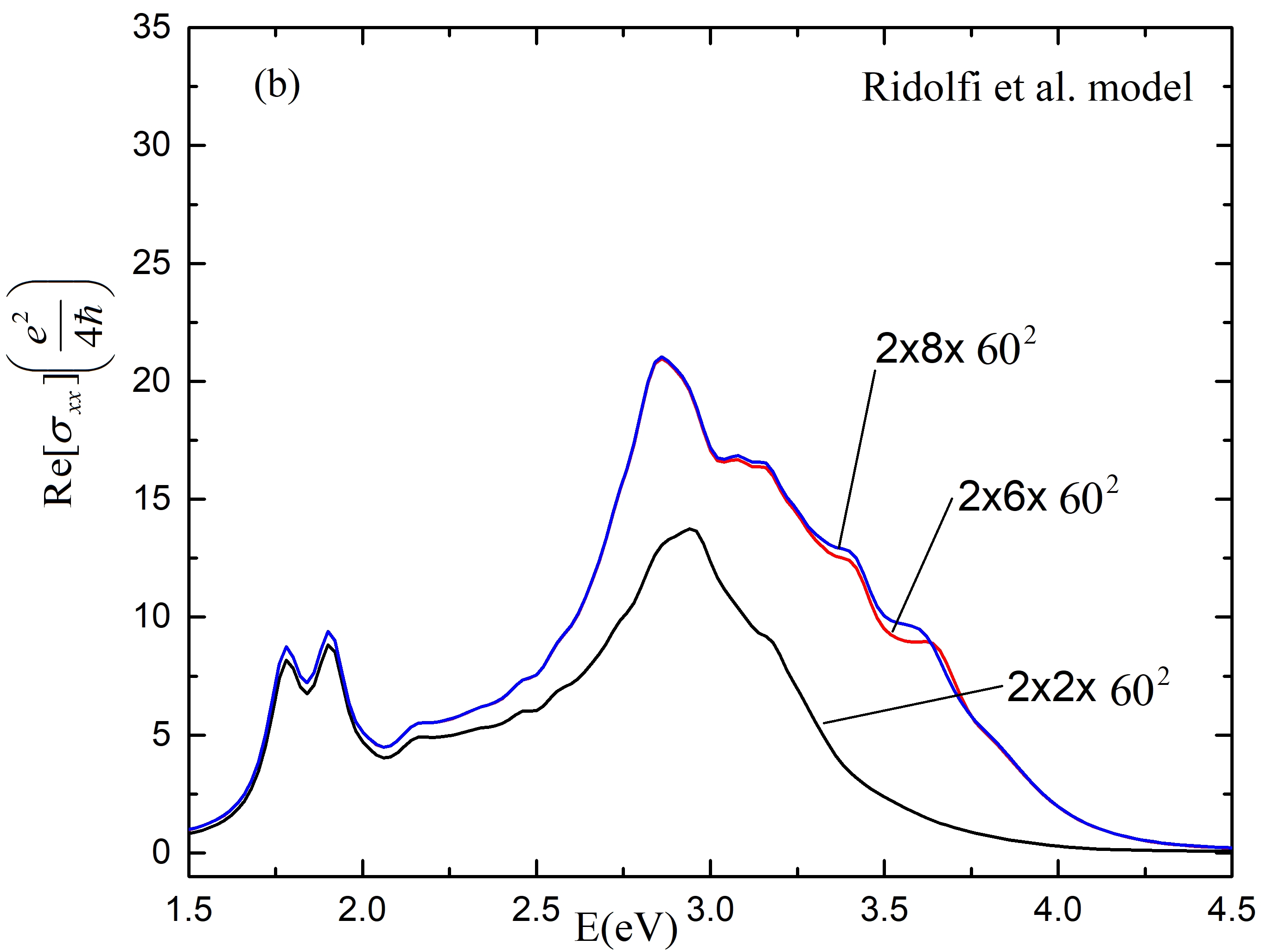}  
\caption{
Linear optical conductivity when considering different numbers of 
conduction bands in the models of Wu \emph{et al.} \cite{Wu2015} (a) and 
Ridolfi \emph{et al.} \cite{Ridolfi2015} (b). 
}
\label{fig:bandssnumber}
\end{figure}

Reference~\onlinecite{Qiu2013} reports that the C peak is contributed by 6 
nearly degenerate exciton states made from transitions between the highest 2 
valence bands and the first three lowest conduction bands (including spin). In 
\Fref{fig:bandssnumber} we calculate $\Re\sigma_{xx}$ for $N_{c}\teq 2$, 
$N_{c}\teq 6$ and  $N_{c}\teq 8$ and verify the necessity of including at least 
$N_{c}\teq 6$ bands. We note that the optical conductivity acquires significant 
corrections due to the increased number of bands precisely in the energy region 
of the C excitons for both TB models. In the model of Wu \emph{et al.} 
\cite{Wu2015}, the C peak is also slightly enhanced when passing from $N_{c}\teq 
6$ to $8$, while in our TB model the most significant changes occur when 
passing from $2$ to $6$. 

\subsection{Even and odd bands}

The mirror symmetry with respect to the horizontal plane that contains the 
transition metal ions has important consequences for the band structure of 
MoS$_2$ monolayers. The TB model that we used to describe the ground-state band 
structure \cite{Ridolfi2015} predicts one even valence band, two even conduction 
bands, and two odd parity conduction bands around the Fermi energy [see 
\Fref{fig:bands}(a)]. The importance of the odd bands has been unclear 
\cite{Qiu2013}, and we examine this issue next.

\begin{figure}[tb]
\centering
\includegraphics[width=0.9\columnwidth]{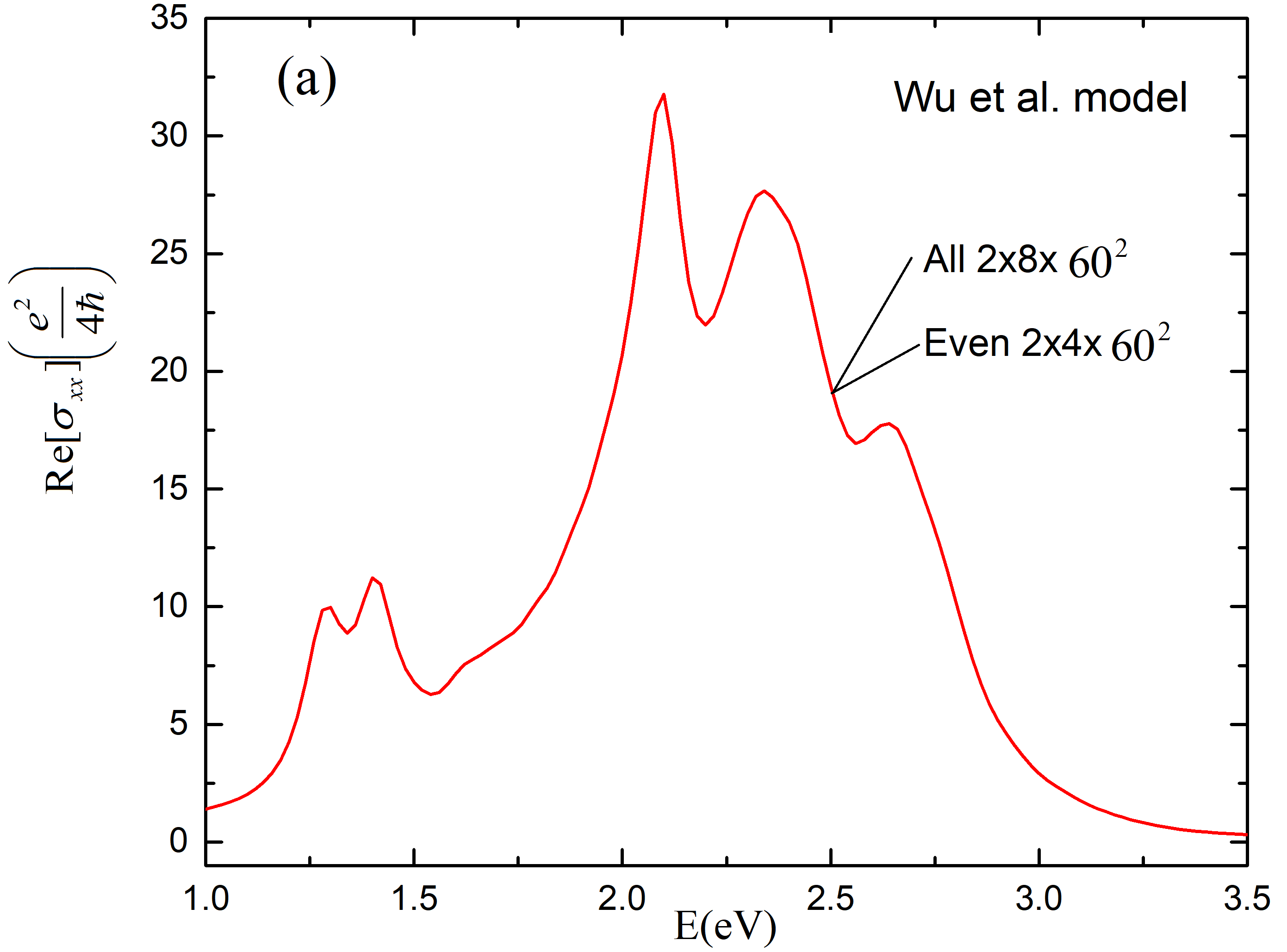}\\
\includegraphics[width=0.9\columnwidth]{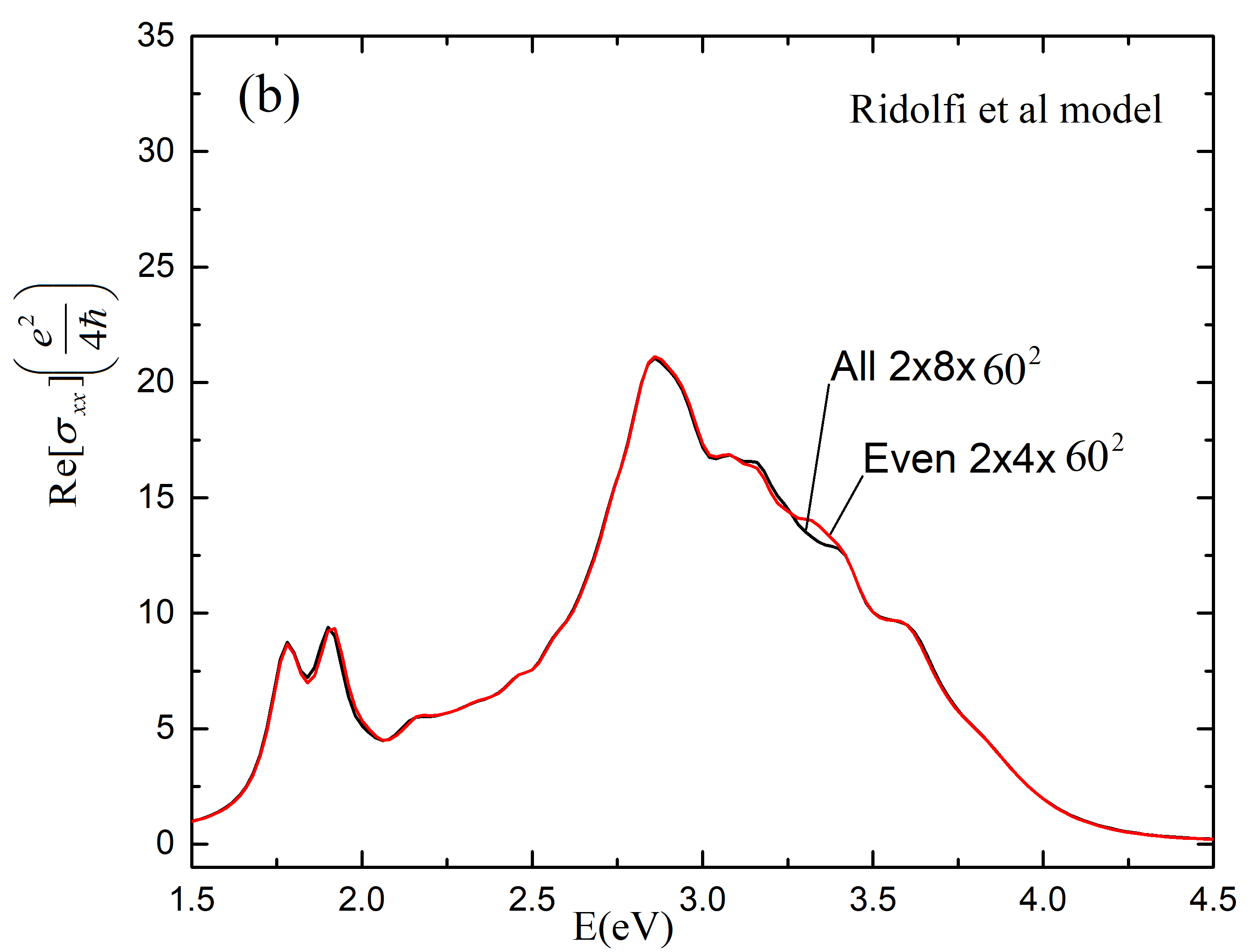} 
\caption{
Comparison between the linear optical conductivity obtained with an 
``all-band'' model ($N_{c}\teq 8$, $N_{v}\teq 2$, $N_{ks}\teq 60$) and an 
``even-band'' 
model ($N_{c}\teq 4$, $N_{v}\teq 2$, $N_{ks}\teq 60$) using the TB 
parameterizations of Wu 
\emph{et al.} \cite{Wu2015} (a) and Ridolfi \emph{et al.} \cite{Ridolfi2015} 
(b).
}
\label{fig:Fanyao_even_vs_all}
\end{figure}

Figure \ref{fig:Fanyao_even_vs_all}(a) shows that the odd bands of the TB model 
of \Ref~\onlinecite{Wu2015} do not contribute to the optical conductivity. 
This is expected because this TB model does not include spin-flipping terms: 
since the dipole coupling is diagonal in spin, the non-zero transition matrix 
elements must involve initial and final states with the same parity under a 
reflection with respect to the plane. Our TB model gives a small difference 
between the all even and odd bands basis caused by the coupling of odd and even 
bands by the spin-flip terms in the Hamiltonian. 
We conclude that the increment in the optical conductivity  when increasing the 
number of bands comes mainly from the two upper ``even'' bands.

\subsection{Exciton and one-electron JDOS}

\begin{figure}[tb]
\centering
\includegraphics[width=0.8\columnwidth]{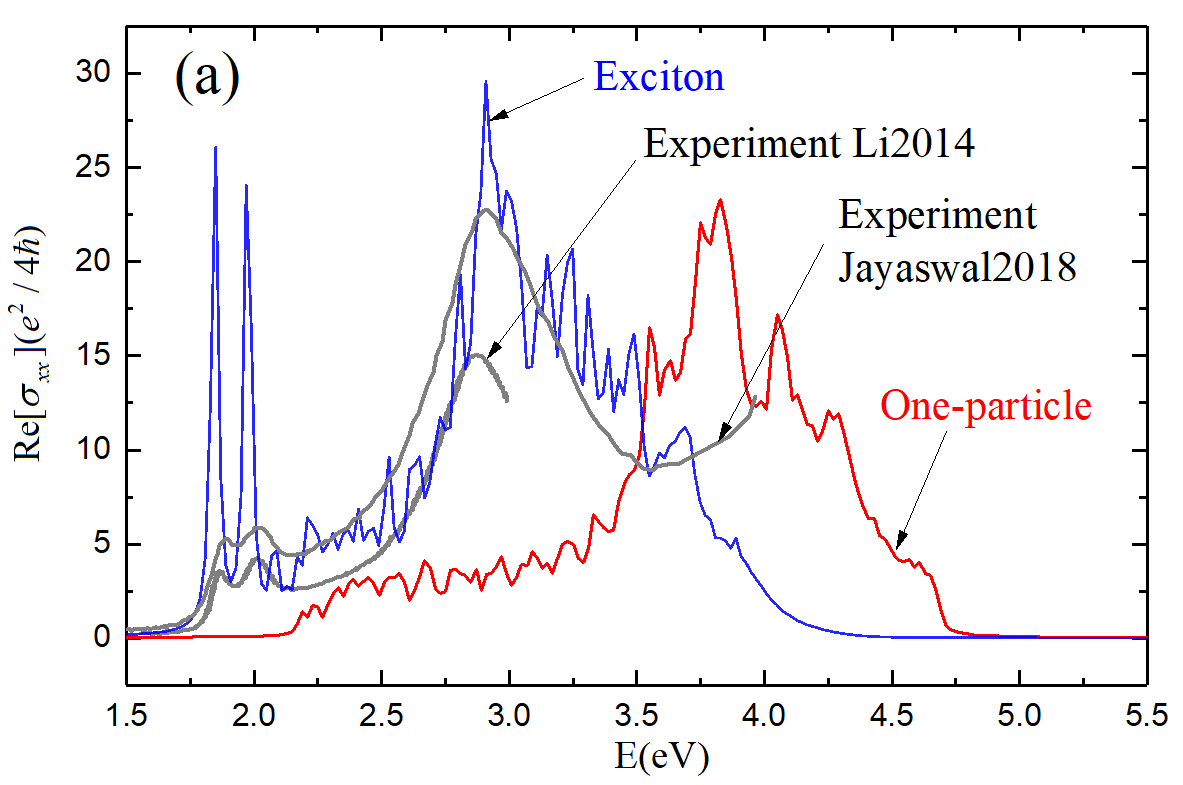}\\
\includegraphics[width=0.8\columnwidth]{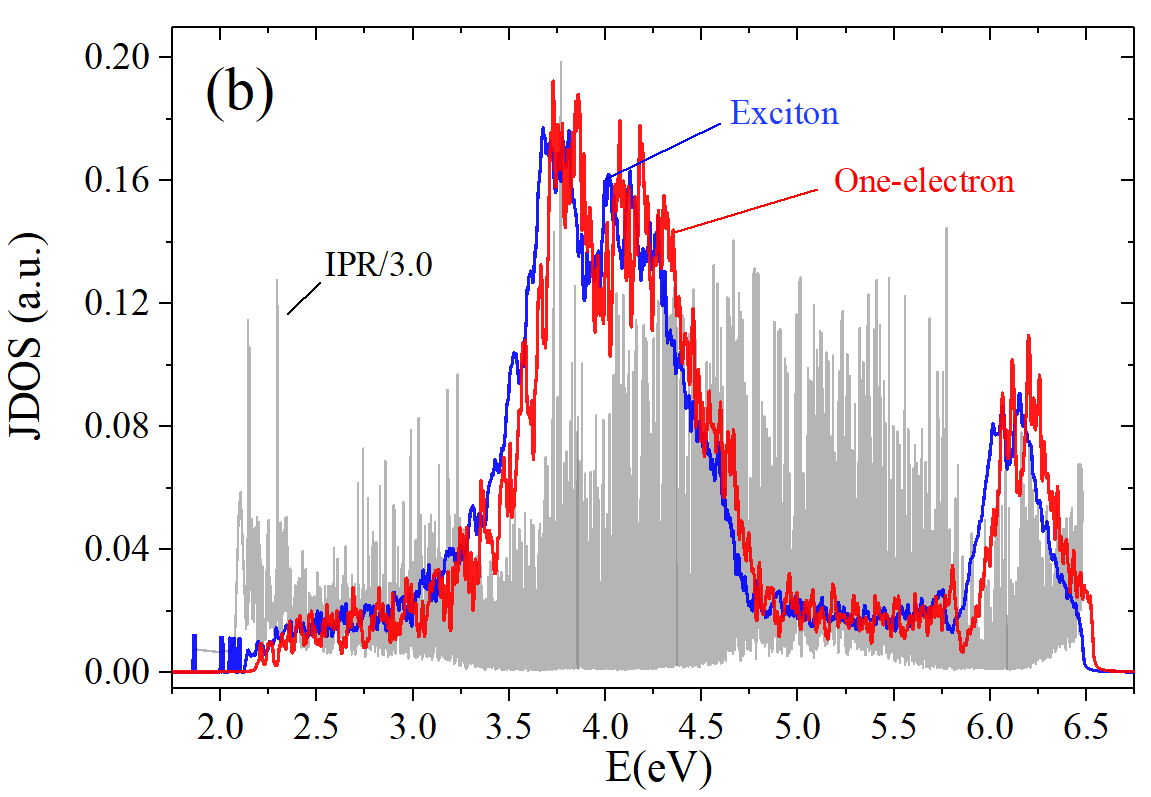}
\caption{
(a) The same as \Fref{fig:sigma}, except that the calculated curves have been 
broadened by the smaller value 0.03\,eV (full width).
(b) Joint density of exciton states (JDOS) as defined in \Eqref{eq:jdos}. A 
Lorentzian level broadening of 0.01\,eV (full width) has been used, except in 
the lower energy range, where we used 0.001\,eV to allow the resolution of 
individual bound exciton levels. The inverse participation ratio (IPR) has been 
calculated from \Eqref{eq:ipr} and is presented without any broadening.
}
\label{fig:sigma-jdos-small-broadening}
\end{figure}

When interpreting the origin of the spectral weight shift in the optical 
conductivity associated with the C excitons, it is useful to analyze how the 
excitation spectrum itself changes in the presence of the Coulomb interaction. 
\Fref{fig:sigma-jdos-small-broadening}(a) shows the same data for 
$\Re\sigma(\omega)$ that was presented in \Fref{fig:sigma}, but using a 
considerably smaller broadening to reveal more clearly the fine spectral 
structure. 
In \Fref{fig:sigma-jdos-small-broadening}(b) we compare the joint density of 
states (JDOS) with and without interaction. Apart from the emergence 
of the bound excitonic states in the gap, we can see that the excitation 
spectrum largely maintains the JDOS computed at the one-electron level. The 
interaction causes a global redshift of about $0.1$\,eV, which is much smaller 
than the spectral weight transfer seen in the conductivity. This figure 
additionally includes the inverse participation ratio (IPR) of all the 
excitonic levels \eqref{eq:ipr}, which quantifies the degree of localization of 
the respective wave functions in reciprocal space.

\subsection{Exciton inverse participation ratio}

To have an overall perspective over the degree of localization of each exciton's 
wave function, we computed the inverse participation ratio (IPR),
\begin{equation}
  \mathcal{P}(E_M) \equiv \sum_{cv\bk} |A^M_{cv\bk}|^4 / \sum_{cv\bk} 
  |A^M_{cv\bk}|^2,
  \label{eq:ipr}
\end{equation}
for all the eigenfunctions of the BSE \eqref{eq:BSE}. This quantity 
provides a rough measure of the spread (in \emph{reciprocal} space) of the 
wave function belonging to the exciton with energy $E_M$, being largest for the 
most localized states and scaling ${\propto\,}1/N_{\text{tot}}$ for states 
uniformly extended over the whole BZ. The result is included in 
\Fref{fig:sigma-jdos-small-broadening}(b). It reveals that while, on the one 
hand, the region of the C excitons is characterized by a comparatively small 
JDOS, on the other, states there are typically more localized than the average, 
as revealed by a number of peaks of the IPR in the interval 
$[2.75,\,3.25]$\,eV. This simply reflects what has been inferred from the 
selected wave functions shown in \Fref{fig:wavefunctions} and, moreover, 
confirms our earlier statement that C excitons are considerably more localized 
than bound ones in reciprocal space: $\mathcal{P}(E\tapprox E^{A,B}) {\,\ll\,} 
\mathcal{P}(E\tapprox E^C)$. This, of course, is as expected because the latter 
are true bound states in \emph{real} space (in relation to this, note that a 
pure Bloch state has $\mathcal{P}(\ve_{\bk}) \teq 1$ since its wave function is 
entirely localized at the point $\bk$ in the BZ).

\bibliography{MoS2_excitons-v24}

\end{document}